\documentclass[
reprint,
superscriptaddress,
amsmath,
amssymb,
longbibliography,twocolumn
]{revtex4-2}

\usepackage{amsmath}
\usepackage{mathtools}
\usepackage{amssymb}
\usepackage{graphicx}
\usepackage{bm}
\usepackage[colorlinks, linkcolor=blue, citecolor=blue, urlcolor=blue,breaklinks=true]{hyperref}
\usepackage{color,dsfont,multirow,booktabs}
\usepackage{braket}
\usepackage{tabularx}
\usepackage{lipsum,booktabs}

\usepackage{ulem}
\newcommand{\pb}[1]{\textcolor{red}{#1}}

\begin{document}

\date{\today}

\title{Selective Single and Double-Mode Quantum Limited Amplifier}

\author{Abdul Mohamed}
\address{Department of Physics and Astronomy, University of Calgary, Calgary, AB T2N 1N4 Canada}
\author{Elham Zohari}
\address{Department of Physics and Astronomy, University of Calgary, Calgary, AB T2N 1N4 Canada}
\address{Department of Physics, University of Alberta, Edmonton, AB T6G 2E1 Canada}
\author{Jarryd J. Pla}
\address{School of Electrical Engineering and Telecommunications, UNSW Sydney, Sydney NSW 2052, Australia}
\author{Paul E.\ Barclay}
\address{Department of Physics and Astronomy, University of Calgary, Calgary, AB T2N 1N4 Canada}
\author{Shabir Barzanjeh}
\address{Department of Physics and Astronomy, University of Calgary, Calgary, AB T2N 1N4 Canada}

\begin{abstract}

A quantum-limited amplifier enables the amplification of weak signals while introducing minimal noise dictated by the principles of quantum mechanics. These amplifiers serve a broad spectrum of applications in quantum computing, including fast and accurate readout of superconducting qubits and spins, as well as various uses in quantum sensing and metrology. Parametric amplification, primarily developed using Josephson junctions, has evolved into the leading technology for highly effective microwave measurements within quantum circuits. Despite their significant contributions, these amplifiers face fundamental limitations, such as their inability to handle high powers, sensitivity to parasitic magnetic fields, and particularly their limitation to operate only at millikelvin temperatures.
To tackle these challenges, here we experimentally develop a novel quantum-limited amplifier based on superconducting kinetic inductance and present an extensive theoretical model to describe this nonlinear coupled-mode system. Our device surpasses the conventional constraints associated with Josephson junction amplifiers by operating at much higher temperatures up to 4.5 K. With two distinct spectral modes and tunability through bias current, this amplifier can operate selectively in both single and double-mode amplification regimes near the quantum noise limit. Utilizing a nonlinear thin film exhibiting kinetic inductance, our device attains gain exceeding 50 dB in a single-mode and 32 dB in a double-mode configuration while adding $0.82$ input-referred quanta of noise. Importantly, this amplifier eliminates the need for Josephson junctions, resulting in significantly higher power handling capabilities than Josephson-based amplifiers. It also demonstrates resilience in the presence of magnetic fields, offers a straightforward design, and enhances reliability. This positions the amplifier as a versatile solution for quantum applications and facilitates its integration into future superconducting quantum computers.
\end{abstract}

\maketitle
\section{Introduction}
Parametric amplifiers with noise characteristics at the quantum limit play a central role in quantum processors \cite{PhysRevD.26.1817, RevModPhys.82.1155}, offering significant applications across various quantum systems \cite{9134828, RevModPhys.93.025005}. In the microwave frequency range \cite{Castellanos, PhysRevB.83.134501, Bergeal2010, JTWPA, JPA_Review, Esposito} these amplifiers enable precise, fast, and high-fidelity single-shot measurements of superconducting qubits \cite{PhysRevApplied.7.054020, Heinsoo2018, PhysRevLett.106.110502, PhysRevLett.112.190504}, ensembles of spins \cite{doi:10.1126/sciadv.adg1593}, quantum dots \cite{PhysRevLett.124.067701, PhysRevApplied.4.014018}, and nanomechanical resonators \cite{PhysRevX.3.021013}. Furthermore, the utilization of near-quantum-limited amplifiers has created new experimental avenues for producing nonclassical radiation, including single and double-mode vacuum squeezing \cite{Castellanos-Beltran2008, PhysRevLett.117.020502, PhysRevLett.106.220502} with applications in weak measurement \cite{Murch2013, Weber2014}, Axion dark matter detection \cite{PhysRevLett.118.091801, PhysRevLett.125.221302, PRXQuantum.2.040350, Backes2021} and quantum illumination and radar \cite{PhysRevLett.114.080503, Barzanjeh,Assouly2023, 2310.07198}. 

The most common method for building microwave parametric amplifiers involves superconducting Josephson junctions, configured as a Josephson Parametric Amplifier (JPA) \cite{PhysRevA.39.2519, Renger2021} or Josephson Traveling Wave Parametric Amplifier (JTWPA) \cite{JTWPA, JPA_Review, Esposito}. These systems exploit the nonlinear characteristics of Josephson junctions to attain significant amplification with minimal noise. Through precise control facilitated by an external microwave pump, JPAs allow for fine-tuning of gain, enabling accurate and efficient amplification of weak signals. Their tunability and instantaneous bandwidth make quantum-limited amplifiers essential tools for advancing quantum information processing and pushing the boundaries of quantum technologies. 

Despite their many valuable applications, both JPAs and JTWPAs encounter significant limitations that restrict their effective integration into scalable quantum architectures. Both JPAs and JTWPAs have a limited saturation input power, primarily attributed to the presence of higher-order nonlinearity such as the Kerr nonlinearity. This limitation typically confines them to operate only with a few microwave photons per bandwidth \cite{Castellanos, PhysRevApplied.11.034014} and thus limits their ability to effectively handle a wide range of frequencies and higher-power signals.
Moreover, Josephson-based amplifiers are highly sensitive to parasitic magnetic fields, which can negatively impact their performance. To mitigate this sensitivity, magnetic field isolation is often necessary. Furthermore, they typically require operation at temperatures below 1 Kelvin \cite{Castellanos, JPA_Review, Esposito}, which can pose practical challenges in applications where higher temperature operation is desired \cite{PhysRevLett.114.080503, Barzanjeh, Assouly2023}.

An alternative approach to overcome these limitations involves the building of quantum-limited amplifiers without the reliance on Josephson junctions by using superconducting films that possess kinetic inductance. This approach utilizes a simplified single-step lithography process and employs thin films of high-temperature kinetic inductance superconductors, such as Niobium Titanium Nitride (NbTiN) \cite{Stannigel, HoEom2012, 10.1063/1.4980102, PhysRevApplied.13.024056, PRXQuantum.2.010302, PhysRevApplied_jarrydPla, PhysRevApplied.21.024011}. The intrinsic kinetic inductance of the NbTiN film introduces the necessary nonlinearity for achieving quantum-limited amplification, thereby eliminating the need to use Josephson junctions. These amplifiers typically use Three-Wave Mixing (3WM) \cite{PhysRevApplied.13.024056, PRXQuantum.2.010302, PhysRevApplied_jarrydPla, PhysRevApplied.21.024011} or Four-Wave Mixing (4WM) \cite{PRXQuantum.4.010322, PhysRevApplied.19.034024} processes to amplify the input signal. The amplification can happen through degenerate or non-degenerate parametric amplification \cite{PhysRevD.26.1817, RevModPhys.82.1155}. In the case of degenerate or phase-preserving amplification, the idler and signal have identical frequencies. This configuration offers the potential for noiseless amplification and the generation of single-mode vacuum squeezing \cite{PhysRevLett.117.020502, Zhong_2013}. In contrast, in non-degenerate or phase-insensitive amplification, the idler and signal exist as separate modes with distinct frequencies. Within this regime, the amplifier's output can be a two-mode squeezed state or an entangled state \cite{RevModPhys.82.1155, Barzanjeh2019, PhysRevLett.128.153603}. Resonance-based kinetic inductance amplifiers primarily operate in the single mode configuration, functioning within degenerate or near non-degenerate amplification regimes \cite{PhysRevApplied_jarrydPla}. In these configurations, shifting the pump frequency from twice the resonance frequency effectively initiates a 3WM process, wherein a pump photon is downconverted to idler and signal photons within the resonator mode. However, this approach encounters limitations, particularly when there is a requirement for generating two-mode squeezing and entanglement. This is primarily because the idler and signal do not represent two spectrally separated modes. Consequently, the development of a quantum-limited amplifier that is tunable, low-noise, compatible with high temperatures, and possesses the ability to selectively operate in both single and two-mode regimes, remains an outstanding challenge.

In this work, we introduce a quantum-limited Kinetic Inductance Parametric Amplifier (KIPA) that overcomes some of the technical limitations associated with Josephson junction amplifiers and is capable of operating at higher temperatures due to the high critical temperature of the kinetic inductance thin film \cite{Annunziata_2010}. The presence of two spectrally and spatially distinct modes, along with the ability to tune the resonance frequency by applying bias current, enables the KIPA to operate selectively in both single and dual-mode amplification regimes. 

By utilizing a uniformly evaporated thin film of NbTiN on high-resistivity intrinsic silicon, we achieve exceptional amplification, surpassing 50 dB gain in single mode and 32 dB gain in double mode configurations, respectively, with a gain-bandwidth product of approximately 30 MHz when operating in the 3WM regime. Importantly, our design eliminates the reliance on Josephson junctions, resulting in a significantly higher 1 dB compression point for the output power when compared to amplifiers based on Josephson technology \cite{parametric_amplification_review}. 
The junction-free design of the KIPA also allows a high degree of resilience to stray magnetic fields or adaptability to experiments requiring strong magnetic fields, reaching up to $2$ T \cite{PhysRevApplied.5.044004, 10.1063/1.4931943, PRXQuantum.4.010322,PhysRevApplied.19.034024}. Additionally, compared to other kinetic inductance amplifiers \cite{PhysRevApplied_jarrydPla}, the design of our KIPA is inherently simple and does not require complex circuitry or filtering to generate amplification. Utilizing kinetic inductance for amplification simplifies the design and fabrication processes to a remarkable degree compared to Josephson-based amplifiers. Moreover, it enhances reliability since the non-linearity of our device is an inherent feature of its geometry, eliminating the need for complicated and delicate fabrication procedures. Additionally, we introduce a theoretical model that describes the dynamics of the system and utilize it to fit the experimental results. 
We note that while the utilization of coupled resonators for amplification has been investigated before, previous studies typically focused on configurations where either both resonators contain nonlinearity or where both resonators shared a nonlinear medium \cite{PhysRevApplied.13.024015, PhysRevApplied.13.024014, PhysRevLett.113.110502}. In contrast, our approach here entails coupling a linear resonator to a nonlinear resonator and harnessing 3WM for amplification in the coupled system.

The paper is organized as follows: In Section \ref{theory}, we establish a comprehensive theoretical model to elucidate single-mode and double-mode amplification based on a coupled mode system. Section \ref{experiment} introduces the experimental setup and the KIPA design. Section \ref{singlemodeAmp} presents the experimental results of single-mode and double-mode amplification. In Section \ref{noise}, we discuss the noise properties of the KIPA and study the operation of the device in different device temperatures. Finally, the concluding remarks and discussions will be presented in Section \ref{conclusion}.

\begin{figure}

    \centering 
     \includegraphics[width=1\linewidth]{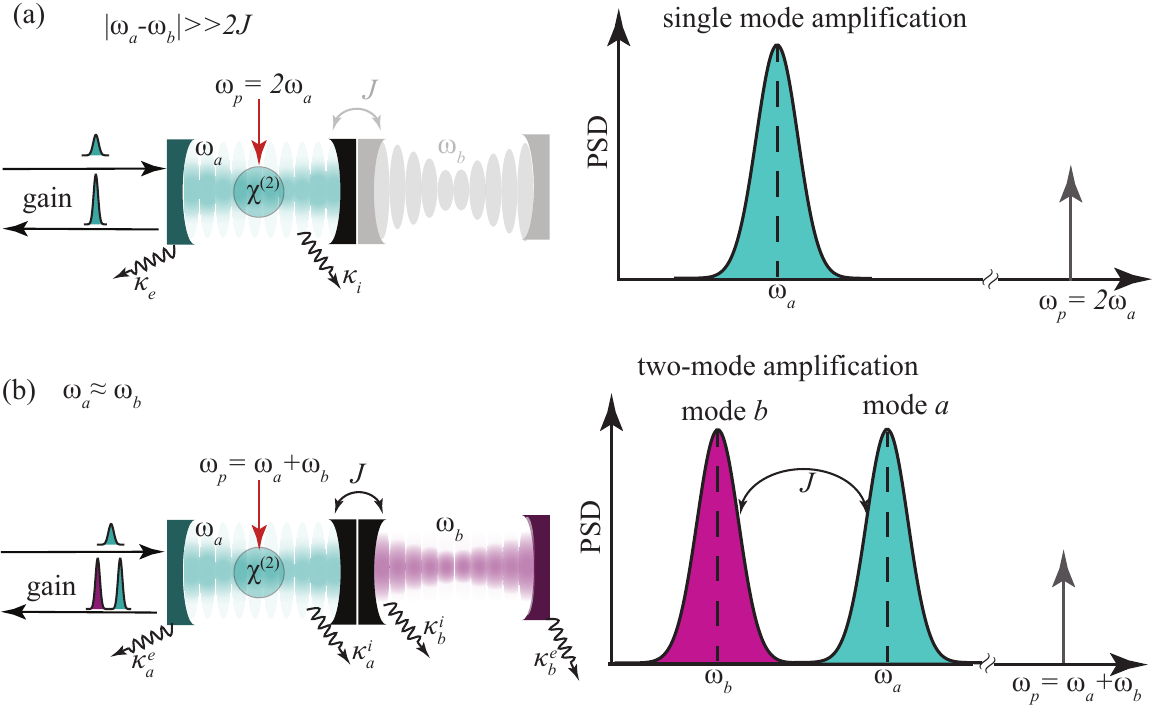}
       \caption{ A schematic representation of a coupled resonator system exhibiting a coupling rate $J$. Here, one resonator, with resonance frequency $\omega_a$ and annihilation operator $a$, is coupled to an auxiliary resonator having resonance frequency $\omega_b$ and annihilation operator $b$. Incorporating a nonlinear medium characterized by a $\chi^{(2)}$ nonlinearity, resonator $a$ facilitates amplification via the Three-Wave-Mixing (3WM) process. The system's operation mode, whether single-mode or double-mode amplification, can be selected by properly pumping the system. (a) Illustrates the single-mode amplification scenario, characterized by a significant separation in resonance frequencies between the two resonators, expressed as $|\omega_a - \omega_b| \gg 2J$. In this setup, the system can be effectively treated as a single mode by disregarding the dynamics of the auxiliary resonator $b$. Amplification of the single mode is achieved by applying a strong pump at $\omega_p = 2\omega_a$, resulting in a degenerate amplification in mode $a$, as seen in the power spectral density (PSD)  of the output field. (b) In contrast, by adjusting the resonance frequencies of the two modes to be equal, i.e., $\omega_a = \omega_b$ and pumping the system at $\omega_p = \omega_a + \omega_b$, double-mode amplification is realized, with the two modes being spectrally separated by the coupling rate $J$. Here, $\kappa_j^{e(i)}$ is the extrinsic (intrinsic) damping rate of each resonator $j=a,b$. }
        \label{Fig1}
\end{figure}

\section{Theoretical modeling of amplification}\label{theory}

In this section, we provide an in-depth theoretical model that describes the 3WM process and amplification within a generic two-mode coupled system. In the subsequent sections, we divide our discussions into single-mode and two-mode amplification, and we construct separate comprehensive theoretical models for each of these regimes.
\subsection{Hamiltonian}
The system under consideration includes a coupled system that incorporates nonlinearity in one of the modes, as shown in Fig. \ref{Fig1}. The Hamiltonian describing this coupled-mode system, accounting for parametric processes and operating in the presence of a pump with a frequency $\omega_p$ and an annihilation operator $a_p$, can be written as (with $\hbar=1$)

\begin{equation} \label{eq:hamiltonian00}
    H=\omega_{a} a^\dagger a+\omega_{b} b^\dagger b+\omega_{p} a_p^\dagger a_p+J(a^\dagger b+b^\dagger a)+H_{\text{non}}. 
\end{equation}
where $a$ and $b$ are the annihilation operators of each mode with frequencies $\omega_a$ and $\omega_b$, respectively, and with coherent mode coupling strength $J$. We also define the nonlinear Hamiltonian \cite{PhysRevApplied.8.054030, PhysRevApplied_jarrydPla}
\begin{equation}
  H_{\text{non}}= \frac{g_0}{2} (a^2a_p^\dagger+a^{\dagger2}a_p)+Ka^{\dagger 2}a^2,
\end{equation}
where the second term in this Hamiltonian describes the degenerate parametric interaction with strength $g_0$ and the third term represents a four-wave mixing interaction with strength $K$. In this Hamiltonian, the inclusion of nonlinearity in one of the modes results in an interesting scenario where we can operate the device in either single or double-mode amplification regimes. This choice can be made by tuning the resonance frequency of mode $a$, either far from or close to the anticrossing point, where efficient exchange of energy between the two modes occurs. We also note that, for our KIPA in the presence of a strong bias current, $K\ll g_0$, and therefore the four-wave mixing process can be neglected. With a film thickness of 10 nm and a resonator of width $2\,\mu$m, the estimated Kerr nonlinearity will be approximately $K \approx 1$ mHz. As we will see later, this value is a few orders of magnitude smaller for the auxiliary resonator $b$. 

\subsection{Single mode amplification}
Away from the anticrossing point, where the frequency of mode $a$ is well-separated from the auxiliary mode $b$, the interaction between these modes does not facilitate photon hopping or exchange of the excitation. Consequently, mode $b$ does not undergo amplification when $|\omega_a- \omega_b|\gg 2J$. In this case, we can treat the system as a single mode, characterized by substantial nonlinearity in mode $a$, see Fig. \ref{Fig1}a. Therefore the Hamiltonian of the system in the strong pump regime is given by (see Supplementary Materials)
\begin{equation} \label{Ham1}
    H=\Delta a^\dagger a+\frac{g}{2} (a^2e^{-i\phi_{p}}+a^{\dagger2}e^{i\phi_{p}}),
\end{equation}
where $g=g_0|\alpha_p|$ and $|\alpha_p|$ is the amplitude of the pump and $\phi_p$ is the global phase set by the pump. The above Hamiltonian has been written in a reference frame rotating at $\omega_p/2=\omega_a-\Delta$.

The system's output can be written with respect to the input quantum noise (\(a_\mathrm{e}\) with an extrinsic damping rate of \(\kappa_\mathrm{e}\)) and the intrinsic losses in the resonator mode (expressed as \(a_\mathrm{i}\) with an intrinsic damping rate of \(\kappa_\mathrm{i}\)). Solving the quantum Langevin equations allows us to obtain the output field of the resonator (see Supplementary Materials)
\begin{eqnarray}\label{aout0}
  a_\mathrm{out}(\omega)&=&\Big[\mathcal{G}_S(\omega)a_\text{e}+\mathcal{G}_I(\omega)a_\text{e}^\dagger\Big]\nonumber\\ 
  &+&\sqrt{\frac{1-\eta}{\eta}}\Big[(\mathcal{G}_S(\omega)+1)a_\text{i}+\mathcal{G}_I(\omega)a_\text{i}^\dagger \Big]  
\end{eqnarray}
 where $\eta=\frac{\kappa_\text{e}}{\kappa}$ describes the waveguide-resonator coupling with $\kappa=\kappa_\text{e}+\kappa_\text{i}$ and we define
 \begin{eqnarray}\label{gainS}
   \mathcal{G}_S(\omega)&=&\frac{\eta \kappa \Big[\frac{\kappa}{2}-i(\omega+\Delta)\Big]}{\Delta^2-g^2+(i\omega-\frac{\kappa}{2})^2}-1,\nonumber\\
   \mathcal{G}_I(\omega)&=&\frac{-i\eta \kappa g e^{i\phi}}{\Delta^2-g^2+(i\omega-\frac{\kappa}{2})^2},
 \end{eqnarray}
Note that $|\mathcal{G}_I(\omega)|^2=|\mathcal{G}_S(\omega)|^2-1$ for $\eta\approx 1$. The first two terms of Eq. (\ref{aout0}) show the amplification of the signal $a_S \equiv a_e$ and idler $a_I\equiv a_e^\dagger$ modes while the last two terms represent the noise terms added by the KIPA due to the intrinsic loss of the resonator. In the absence of the internal loss $\eta \rightarrow 1$, Eq. (\ref{aout0}) simplifies to $a_\mathrm{out}(\omega)=\sqrt{G}a_\text{e}+\sqrt{G-1}a_\text{e}^\dagger$, describing the annihilation operator of a degenerate (phase-sensitive) parametric amplifier with gain
\begin{equation}
G=|\mathcal{G}_S|^{2}=\Big(\frac{2}{1-4(g/\kappa)^2}-1\Big)^2   
\end{equation}
here for simplicity, we consider $\Delta=\omega=0$. We can see that for $g\ll \kappa$ the gain is negligible $G\rightarrow 1$. However, the highest level of gain is achieved near the denominator's roots and as $g\rightarrow \kappa/2$ which results in $G\rightarrow \infty$. This divergence can be understood better by looking at the stability condition of the system through the susceptibility of the resonator \cite{Levitan_2016}. For $\Delta=0$, the resonator's susceptibility is given by (see the Supplementary Materials)
\begin{equation}
   \boldsymbol{\chi}(\omega)=\frac{1}{(i\omega-\frac{\kappa}{2})^2-g^2}\begin{bmatrix}
-i\omega+\frac{\kappa}{2} & -ig e^{i\phi_p} \\
ig e^{-i\phi_p} & -i\omega+\frac{\kappa}{2}
\end{bmatrix}
\end{equation}
The stability of the system is determined by the denominator of $\boldsymbol{\chi} (0)$, imposing $g<\kappa/2$ as the stability criteria  \cite{RevModPhys.82.1155}. This condition ensures that the parametric amplifier operates below its threshold and avoids entering self-sustained oscillations. However, when $g\geq \kappa/2$ the system becomes unstable and it leads to parametric self-oscillations.

\subsection{Double-mode amplification}
The result in Eq. (\ref{aout0}) can be extended to non-degenerate, phase-insensitive amplification, where we deal with two spectrally distinct signal and idler modes. 
This can be achieved by bringing mode $a$ on resonance with mode $b$, see Fig \ref{Fig1}b. The coherent interaction between the two modes, when combined with the 3WM process in mode $a$, enables the non-degenerate amplification within the coupled system. This becomes evident when we diagonalize the interaction Hamiltonian and write Eq. (\ref{eq:hamiltonian00}) in the hybridized (collective) modes representation $c_{\pm}=\frac{a\pm b}{\sqrt{2}}$, reads
\begin{eqnarray} \label{TMHam}
H=\sum_{i=\pm}\omega_\pm{}c_{i}^{\dagger}c_{i}+\omega_{p} a_p^\dagger a_p&+&\frac{g_0}{2}\Big[c_{+}c_{-}a_p^\dagger+\text{h.c}\Big] \\
&+&\frac{g_0}{4}\Big[\big(c_{+}^{2}+c_{-}^{2}\big)a_p^\dagger+\text{h.c}\Big].\nonumber
\end{eqnarray}
where $\omega_{\pm}=\frac{\omega_a+\omega_b}{2}\pm \sqrt{\frac{\Delta_{ab}^2}{4}+J^2}$ are the frequencies of the collective modes and $\Delta_{ab}=\frac{\omega_a-\omega_b}{2}$. In the strong pump regime and a reference frame rotating at $\omega_p/2=\Omega-\Delta$, the Hamiltonian (\ref{TMHam}) reduces to
\begin{eqnarray} \label{eq:hamiltonian}
H&=&\Delta_+c_{+}^{\dagger}c_{+}+\Delta_- c_{-}^{\dagger}c_{-}\\
&+&\frac{g}{2}\Big[e^{-i\phi_p}\big(c_{+}^{2}+c_{-}^{2}\big)+\text{h.c}\Big]+g\Big[e^{-i\phi_p}c_{+}c_{-}+\text{h.c}\Big]. \nonumber
\end{eqnarray}
where $g=g_0|\alpha_p|$ and we consider $\Omega\equiv\omega_a=\omega_b$. Note that the third term in this Hamiltonian describes the degenerate amplification in the collective modes, while the last term represents the non-degenerate amplification. The presence of the terms $\Delta_{\pm}=\Delta\pm J$ in the first two parts of the Hamiltonian enables us to effectively select (activate) degenerate or non-degenerate amplification, given that $2J$ exceeds the total damping rates of the modes $\kappa_{\pm}$ and the coupling rate $g$. This condition becomes evident when we move to the interaction picture with respect to $\Delta_+ c_{+}^{\dagger}c_{+}+\Delta_-c_{-}^{\dagger}c_{-}$, 
\begin{eqnarray} \label{eq:hamiltonianint}
H_\text{int}&=&\frac{g}{2}\Big[e^{-i\phi_p}\big(c_{+}^{2}e^{-2i\Delta_+t}+c_{-}^{2}e^{-2i\Delta_-t}\big)+\text{h.c}\Big]\nonumber\\
&+&g\Big[e^{-i\phi_p}c_{+}c_{-}e^{-2i\Delta t}+\text{h.c}\Big]. 
\end{eqnarray}

By choosing $\Delta=J$ and under Rotating Wave Approximation (RWA), we effectively disregard the rapidly oscillating terms rotating at a frequency of $2J$. This choice enables us to specifically select degenerate amplification in mode $c_-$ while $\Delta=-J$ results in amplification in mode $c_+$. In these situations, the gain in the degenerate parametric amplification can still be described by Eqs. (\ref{aout0}) and (\ref{gainS}) by replacing $a_\text{e/i}\rightarrow c^\text{e/i}_{\pm}$ where $c^\text{e/i}_{\pm}$ are the effective noise operators of the hybridized modes, $\kappa\rightarrow \kappa_{\pm}$, $\Delta\rightarrow \Delta_{\pm}$, and $\eta \rightarrow \eta_{\pm}$. On the other hand, if we choose $\Delta=0$, which corresponds to $\omega_p=\omega_a+\omega_b$, we can activate non-degenerate amplification or select the terms involving $c_{+}c_{-}$ in the Hamiltonian. In this situation, the contributions of single-mode amplification terms proportional to $c_{\pm}^2e^{\mp2iJ}+\text{h.c}$ are negligible under RWA. By solving the quantum Langevin equation we can extract the output field for the hybridized modes
\begin{eqnarray}
  c_\mathrm{\pm,out}(\omega)&=&\mathcal{G}^{\pm}_S(\omega)c_{\pm}^e+\mathcal{G}_I^{\pm}(\omega)c_{\mp}^{e\dagger}\\
  &+&\sqrt{\frac{1-\eta^{\pm}}{\eta^{\pm}}}\Big[(\mathcal{G}^{\pm}_S(\omega)+1)c_{\pm}^i+\mathcal{G}^{\pm}_I(\omega)c_{\mp}^{i\dagger} \Big].\nonumber 
\end{eqnarray}
where we define the gain factor 
  \begin{eqnarray}
  \mathcal{G}^{\pm}_S(\omega)&=&\frac{\eta \kappa_{\pm} \Big[\frac{\kappa_{\mp}}{2}-i(\omega+\Delta_{\mp})\Big]}{[i(\omega-\Delta_+)-\frac{\kappa_+}{2}][i(\omega+\Delta_-)-\frac{\kappa_-}{2}]-g^2}-1, \nonumber\\
\mathcal{G}_I^\pm(\omega)&=&\frac{-i\sqrt{\eta_-\eta_+} \kappa_\pm g e^{i\phi}}{[i(\omega-\Delta_+)-\frac{\kappa_+}{2}][i(\omega+\Delta_-)-\frac{\kappa_-}{2}]-g^2}.\nonumber\\
\end{eqnarray}
and $|\mathcal{G}^{\pm}_I(\omega)|^2=|\mathcal{G}^{\pm}_S(\omega)|^2-1$ for $\eta=1$.
The stability condition for the double-mode amplification can be extracted based on the damping parameters of the initial (non-hybridized) modes, leading to $g<\kappa_a(1+\mathcal{C}_0)/2$ where $\mathcal{C}_0=4J^2/(\kappa_a \kappa_b)$ is the cooperatively of the interaction between the two resonators. 
In what follows, we will conduct experimental investigations into both degenerate and non-degenerate amplification within a kinetic inductance superconducting resonator. 
\begin{figure*}[t]
    \centering 
     \includegraphics[width=1\linewidth]{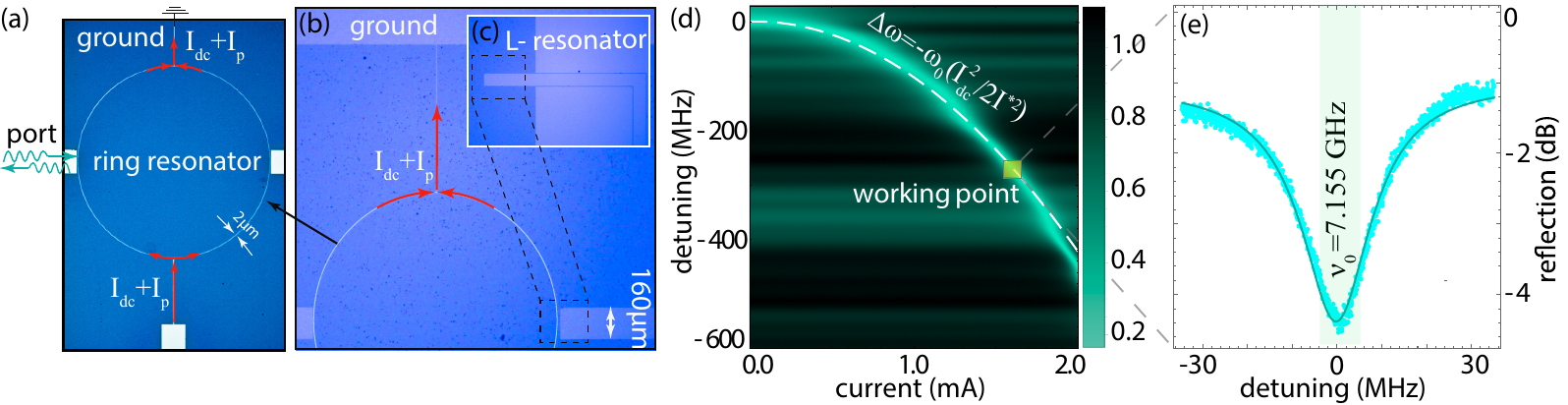}
       \caption{ (a) A ring resonator with a $1.3$ mm diameter and a $2 \,\mu$m pitch is connected capacitively to an auxiliary resonator. The system is probed through a transmission line connected capacitively to the left side of the ring. A separate line provides the pump and DC current from the device's bottom, flowing through the ring. (b) Optical image of the KIPA involving a ring and L-shaped auxiliary resonators. (c) Displays the L-shaped auxiliary resonator with a width of  $160 \,\mu$m and a length of 4 mm. The design prohibits DC current from passing this resonator, effectively setting the 3WM nonlinearity to zero for the auxiliary resonator. (d) The resonance frequency of the ring resonator as a function of applied bias current. The resonance frequency shows a quadratic pattern with respect to the applied current $\Delta \omega=-\frac{\omega_0}{2} \Big(\frac{I_\text{dc}}{I_*}\Big)^2$. (e) The resonance frequency of the resonator at fixed current $I_{\text {dc}}=1.575$ mA corresponding to $\omega_0/2\pi=7.155$ GHz.}
        \label{Fig2}
\end{figure*}

\section{Experimental realization}\label{experiment}
The amplification process relies on the kinetic inductance exhibited by the superconducting film, which displays a nonlinear dependency on the applied current. This nonlinear behavior can be described by the Ginzburg-Landau theory and the total inductance of the system \cite{Pippard1, Pippard2, Annunziata_2010, Zmuidzinas} $L_k(I)=L_0[1+(\frac{I}{I^*})^2]$ where $L_0$ denotes the kinetic inductance of the film in the absence of current (see Supplementary Materials), while $I^*$ is proportional to the critical current of the film and serves as a quantitative gauge of the film's responsiveness to the applied current $I$. 

As shown in Fig. \ref{Fig2}a and b, at the core of this amplifier is a ring resonator, used as a nonlinear and dispersive mixing element. The fabrication process involves a straightforward single optical lithography step, as explained in the Supplementary Materials. The resonator is grounded and galvanically connected to both the pump and DC lines. The generated or amplified signal is directed into a waveguide, which is coupled capacitively to the ring resonator. This design simplifies the measurement process and eliminates the requirement for a Bragg mirror or impedance-matching step coupler \cite{Stannigel, HoEom2012, 10.1063/1.4980102, PhysRevApplied.13.024056, PRXQuantum.2.010302} within the system, resulting in a design that is exceptionally simple and compact. The mixing interaction between idler and signal modes is facilitated by applying a DC current $I_\text{dc}$, modifying the circuit inductance
\begin{equation}
    L_k(I)=L_0\Big[1+\Big(\frac{I_\text{dc}}{I_*}\Big)^2+\frac{2I_\text{rf}I_\text{dc}}{I_*^2}+\Big(\frac{I_\text{rf}}{I_*}\Big)^2\Big],
    \label{Kin}
\end{equation}
where we consider $I=I_\text{dc}+I_\text{rf}$ in which $I_\text{rf}$ is the current of the microwave signal. The first term in Eq. (\ref{Kin}) gives the resonance frequency at zero current, whereas the second term characterizes a shift in the resonance frequency of the resonator due to the DC current $\Delta \omega=-\frac{\omega_0}{2} \Big(\frac{I_\text{dc}}{I_*}\Big)^2$. The third term introduces the 3WM process, and finally, the fourth term leads to Kerr nonlinearity \cite{PhysRevApplied_jarrydPla}. The dependency of the resonance frequency of the mode on the current allows us to extract $I^*$ and therefore back out the critical current of the superconducting sheet. Note that the ring resonator in our setup corresponds to resonator $a$ as shown in Figure \ref{Fig1}, and it possesses a $\chi^{(2)}$ nonlinearity. 

\begin{figure*}[t]
    \centering 
     \includegraphics[width=1\textwidth]{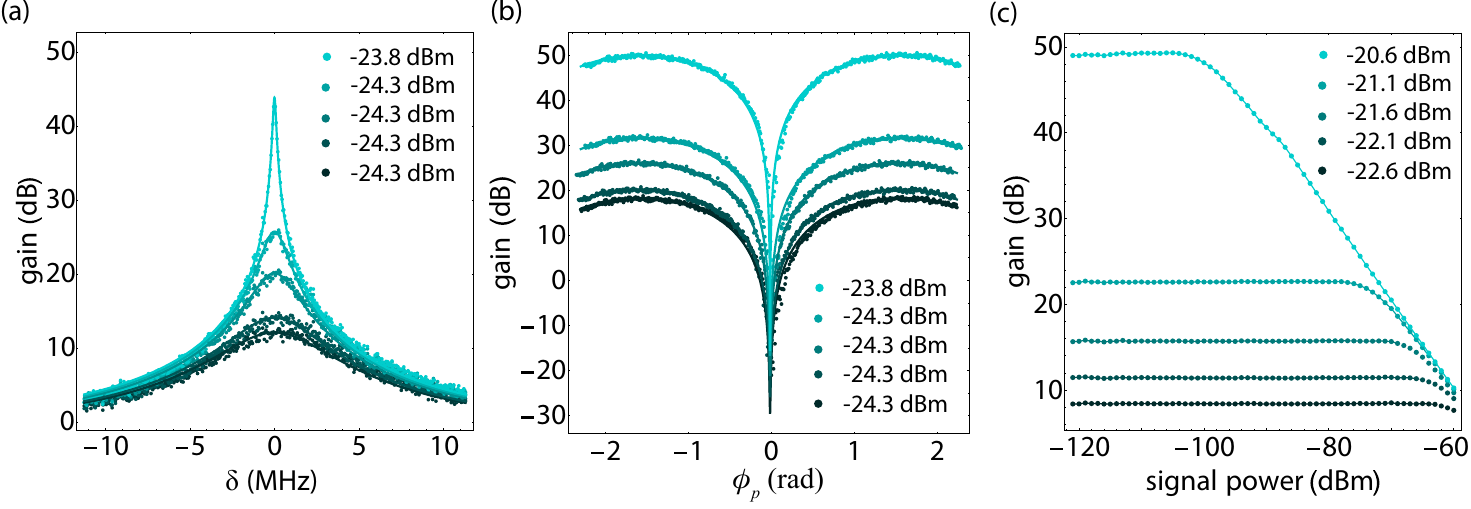}
       \caption{ (a) Single-mode phase-insensitive signal gain versus the frequency detuning from the resonator $\omega=\delta+\frac{\omega_{p}}{2}$ for different pump powers. The experimental results (dots) align well with the theoretical model (solid lines) outlined in the text. (b) Phase-sensitive gain as a function of the pump phase $\phi_p$. A coherent tone positioned at $\delta=0$ is measured while the pump phase is varied. The data is adjusted to align all dips at zero radians. The result is in good agreement with the theory. (c) The 1-dB compression point for a coherent tone at $\delta=1$ kHz from the amplification center frequency. Power is varied to determine the 1dB compression points at different pump powers.}
        \label{Fig4}
\end{figure*}

In addition to the ring resonator, our system also has an auxiliary L-shape resonator (corresponding to resonator $b$ in Fig. \ref{Fig1}), which is capacitively coupled to the ring resonator, see Fig. \ref{Fig2}b and c. The initial intent behind this setup was to have an additional transmission line for dual-sided probing/measuring of the ring resonator.  However, this L-shape section can also act as a resonator with a resonance mode $\omega_b$ close to that of the ring resonator $\omega_a$. As explained in the Supplementary Materials, the behavior of the L-resonator depends on the functionality of a cold switch directly connected to the resonator. When the switch is disconnected from the sample, the L-resonator is essentially floating from both sides. Conversely, when the switch is connected to the sample, the L-resonator is directly coupled to the input transmission line, effectively resulting in a resonator with a significantly large extrinsic coupling rate. Figure. \ref{Fig2} d shows the reflection response of the system when the cold switch is connected to the sample. 

The large size of this auxiliary resonator minimizes its Kerr nonlinearity and allows the handling of substantial power without surpassing the critical current threshold. Moreover, it can be directly wire-bonded, facilitating its utilization as a resonator with significant extrinsic coupling. This capability enables the measurement of the system from an alternative port. The auxiliary resonator is not connected to any DC wires, preventing the flow of biased current through it, thereby maintaining a constant resonance frequency. This setup provides a tunable nonlinear resonator (the ring resonator) connected to a nearby linear auxiliary resonator. This configuration accommodates a pair of coupled modes, initially separated by a spectral gap of approximately $\omega_a-\omega_b \approx360$ MHz at zero bias current, as shown in Fig. \ref{Fig5}a. By applying a DC current, the frequency of the ring resonator can be adjusted, facilitating the alignment of the two modes at the anticrossing point with an intra-mode coupling strength of $J\approx 21.5$ MHz. As described in the theory section, this coupling plays a critical role in selecting either single or double-mode amplification when the resonances reach the anticrossing point.

To initiate the 3WM process within the system, it is necessary to apply a bias current by utilizing a stable and low noise current source to supply the required current $I_{\mathrm{dc}}$. Subsequently, at the millikelvin stage, this current is combined with the microwave pump, represented as $I_p$, through the use of a bias tee. Figure \ref{Fig2}d visually demonstrates the influence of the DC current, showing that an increase in the current induces a downward shift in the frequency of the ring resonator that follows the expression $\Delta \omega=-\frac{\omega_0}{2} \Big(\frac{I_\text{dc}}{I^*}\Big)^2$. By precisely knowing the applied current and fitting the frequency shift, we can deduce that $I^*=5.86$ mA, a value consistent with previously measured transmission lines of the same width \cite{Hortensius_Driessen_Klapwijk_Berggren_Clem_2012} and critical currents of $I_c\lesssim 3.91$ mA\cite{PhysRevApplied_jarrydPla}. Figure \ref{Fig2}e shows the system's reflection from the probe port, demonstrating the resonance frequency of the ring resonator at $\omega_0/2\pi=7.155$ GHz for $I_{\text {dc}}=1.575$ mA. Fitting the resonator mode lineshape provides $\kappa_{\text{e}}=19$ MHz and $\kappa_{\text{i}}=4$ MHz, leading to a waveguide-resonator coupling efficiency of $\eta=0.82$. The second mode is not immediately visible from this figure. However, as described in Fig. \ref{Fig5}a, upon closer examination or measuring with smaller steps, we can distinctly identify the anticrossing point and the existence of mode $b$.

\section{amplification}\label{singlemodeAmp}

\subsection{Single-mode amplification}

To observe single-mode amplification we choose a bias current $I_{\text{dc}}=1.575$ mA far from the anticrossing point. As a result, we can effectively treat the system as single-mode and Hamiltonian (\ref{Ham1}) with $g=-I_\text{dc} I_p \omega_0/(4I^{*2})$ \cite{PhysRevApplied_jarrydPla} can fully describe the amplification process. In this case, the auxiliary resonator will not experience any amplification. We apply a strong pump at a frequency of $\omega_p\approx 14.29$ GHz in addition to the bias current $I_\text{dc}$ which together results in the down-conversion of pump photons, generating signal and idler photons at $\omega_{p}/2$. Figure \ref{Fig4}a shows the non-degenerate gain profile as a function of detuning $\omega=\delta+\frac{\omega_{p}}{2}$ that is measured by sweeping a coherent tone and capturing the maximum gain near half the pump frequency. We observe a substantial gain of approximately 43 dB, achieved with a pump power of $P_p=-23.8$ dBm applied at the device's input. The gain profile is further analyzed using Eq. (\ref{gainS}), resulting in $\kappa_\mathrm{e}=28$ MHz,  $\kappa_\mathrm{i}=4$ MHz, and the coupling efficiency $\eta=0.9$. A comparison of coupling rates in the presence and absence of pump current shows an increase in the extrinsic coupling rate \cite{PhysRevApplied_jarrydPla}, attributed to kinetic inductance changes induced by the presence of the pump.

Additional amplification can be attained by operating the KIPA in a degenerate mode, wherein both the signal and idler have the same frequency, $\delta=0$, and exhibit interference based on the phase difference between the pump and the probe $\Delta \phi$. We use a weak probe signal at precisely half the pump frequency and measure the reflection of the system for various phases of the pump, as shown in Fig. \ref{Fig4}b. The collected data has been shifted to match $\Delta \phi=0$, where we observe a performance range spanning from $-30$ dB of de-amplification and exceeding $50$ dB of amplification. Employing Eq. (\ref{gainS}), we once again fit the data, this time centered at $\omega = 0$. This analysis also provides insights into the extent of squeezing achievable, as during this operation, it becomes possible to squeeze vacuum fluctuations of one quadrature below the standard quantum limit \cite{4WM_Squeeze_Slusher_Hollberg_Yurke_Mertz_Valley_1985}.

\begin{figure*}[t]
    \centering 
     \includegraphics[width=1\textwidth]{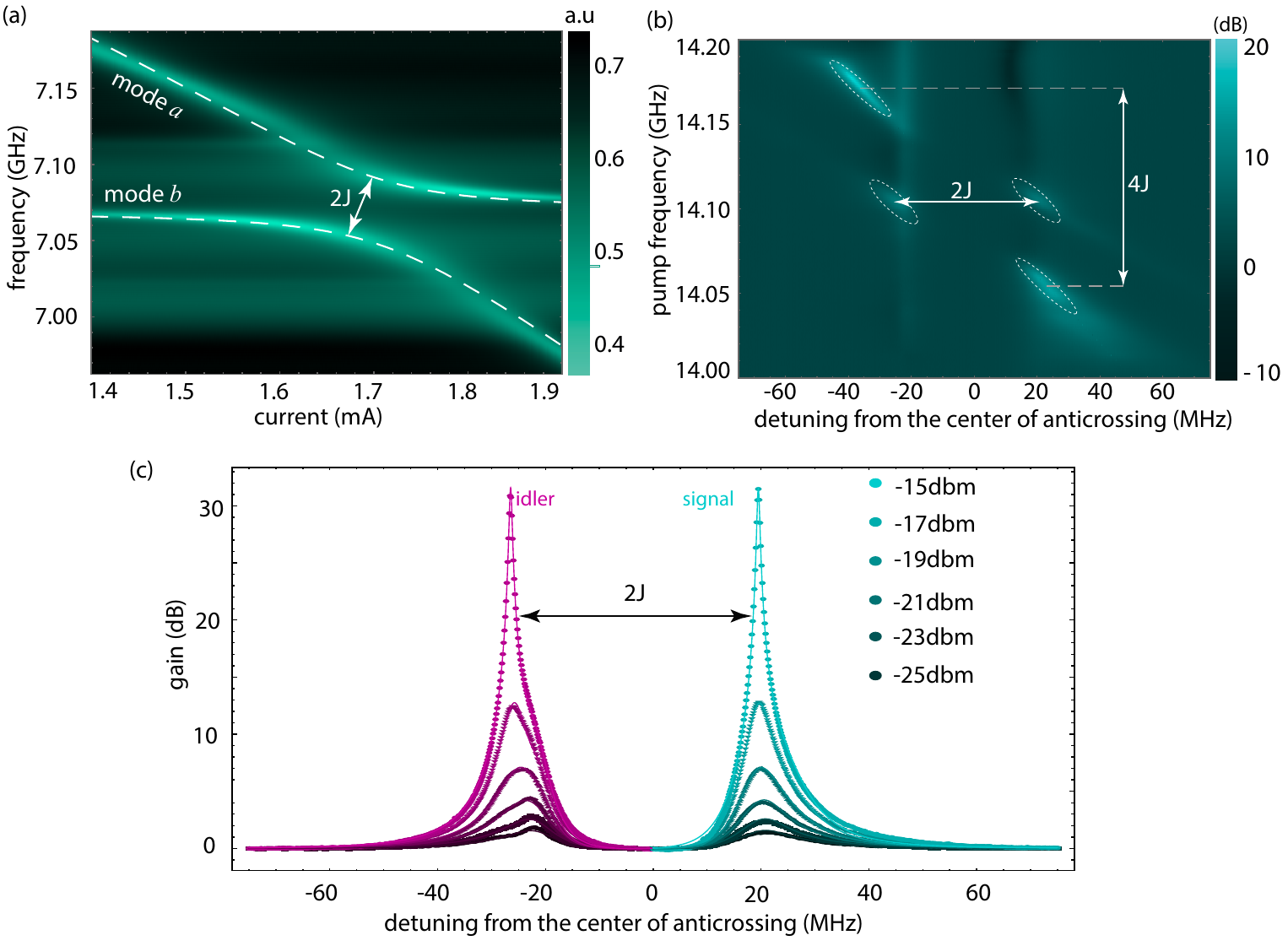}
       \caption{ (a) The appearance of the two modes and the anticrossing point as a function of the applied DC current. The two modes are separated by $2J\approx 43$ MHz. (b) Signal amplification versus pump frequency and detuning from the center of the anticrossing. (c) Phase-insensitive gain for the two modes at various pump powers versus the detuning from the center of the anticrossing. }
       \label{Fig5}
\end{figure*}

Subsequently, we conduct an assessment of the KIPA's 1-dB compression point across various pump power levels. This measurement serves to quantify the maximum input power the amplifier can accommodate before reaching saturation. Once again, we apply a probe signal and determine the point at which the KIPA experiences a 1 dB reduction in its maximum gain by increasing the probe power. The result shows a 1-dB compression power of approximately $-73$ dBm, corresponding to $20$ dB of gain. We note that this compression power is, on average, three orders of magnitude greater than that observed in Josephson-based amplifiers \cite{parametric_amplification_review}.

\subsection{Double-mode amplification}
As previously discussed, the amplifier design studied here accommodates two modes with an initial spectral separation of approximately $360$ MHz at zero bias current. Nevertheless, by applying a DC current to the ring resonator, we can adjust the resonance frequency and shift it toward the anticrossing points. Note that the appearance of nonlinearity and, subsequently, amplification only happens in the ring resonator. This double-mode system can be effectively described using Eq. (\ref{eq:hamiltonian00}), where the nonlinear term $a^2 + a^{\dagger2}$ is only present in the ring resonator due to applying the biased current and mode $b$ describes the auxiliary resonator (L-shape resonator). Figure \ref{Fig5}a shows how the ring resonator, mode $a$, evolves as the bias current is varied, eventually reaching the anticrossing point with the auxiliary resonator with mode $b$. 

By aligning the resonance of the two modes and holding them at the anticrossing point, which is achieved by setting $I_\text{dc}=1.68$ mA, the system can be operated within the non-degenerate (phase-insensitive) amplification regime described by Hamiltonian (\ref{eq:hamiltonianint}). Figure \ref{Fig5}b shows the amplification versus the detuning from the anticrossing point and the pump frequency $\omega_p$. There are three distinct amplification regimes evident in this figure. In the initial region, when $\Delta\approx -J$, single-mode amplification is observed around $\omega_p\approx 14.059$ GHz, corresponding to the term $c^2_-$ in Hamiltonian (\ref{eq:hamiltonianint}). The second region indicates double-mode amplification or the activation of $c_- c_+$ terms at $\Delta \approx 0$,  which is represented by the presence of two peaks at $\omega_p\approx 14.1$ GHz well-separated by $2J$. In the third regime, amplification at $\omega_p\approx 14.15$ GHz occurs when $\Delta\approx J$, corresponding to the term $c^2_+$.  Note that, the pump frequency separation between the first and last regimes, $\delta_p \approx 91$ MHz, is determined by the coupling strength between the two modes $\delta_p \approx 4J$, as anticipated. Figure \ref{Fig5} c demonstrates double-mode (non-degenerate) amplification for different pump powers at $\Delta\approx0$. It shows the amplification of the signal and idler modes achieving up to $32$ dB gain in both modes. The experimental results align well with our theoretical model for the coupled mode system presented in the Supplementary Materials.

\begin{figure*}[t]
     \includegraphics[width=\linewidth]{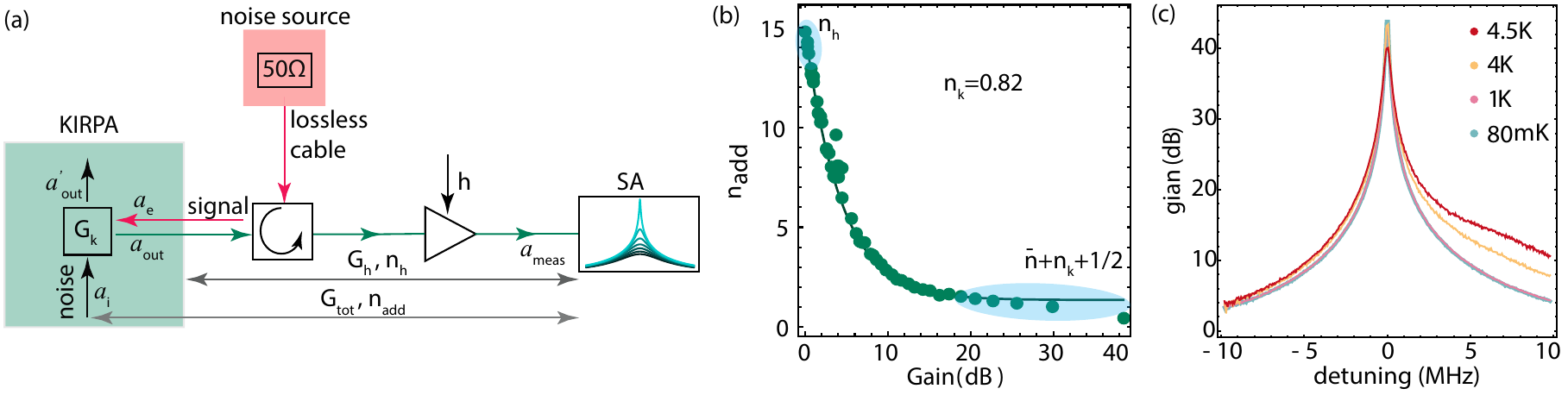}
       \caption{ (a) The illustration of the noise calibration setup. A $50\,\Omega$-terminator, varied in temperature, acts as the known noise source. The generated noise is directed to the KIPA at varying gains. By utilizing the Bose-Einstein distribution to fit the noise spectrum, we extract the total gain and quanta noise added by the amplification chain. (b) The noise quanta added by the whole amplification chain versus the gain of the KIPA when operating at the single-mode regime. At zero gain ($G_\text{k}=0$), the noise of the classical amplification chain $n_\text h$ dominates the added noise. In contrast, at large gains ($G_\text{k}\gg 1$), the noise added by the KIPA becomes noticeable $n_\text{add}\approx (\bar n+\frac{1}{2}+n_{\text{k}})$. Fitting the data using the theoretical model results in $n_\text{k}\approx 0.82$. (c) A comparison of the KIPA gain at different device temperatures ($80$ mK, $1$ K, $4$ K, and $4.5$ K) is presented. }
        \label{Fig6}
  
\end{figure*}
\section{Noise Characterization and Effect of Operating Temperature} \label{noise}
Another important aspect of a parametric amplifier is its noise behavior. We examine the noise performance of the KIPA in the single-mode regime by evaluating a chain of amplifiers and determining the additional noise introduced by the KIPA, as illustrated in Fig. \ref{Fig6}a. The output chain is divided into two parts: the KIPA, with gain $G_\text{k}$ and input-referred added noise $n_\text{k}$, and the classical amplification chain consisting of gain $G_\text{h}$ and noise quanta $n_\text{h}$. This classical chain represents the collective gain and noise arising from the HEMT (near 4K), the low noise amplifier (at room temperature), and the cable losses across the chain. Near the non-degenerate amplification regime, the field operator describing the entire chain is given by

\begin{equation}
a_\text{meas}=\sqrt{G_\text{h}}a_\text{out}+\sqrt{G_\text{h}-1}h^\dagger,
\end{equation}
where $h$ represents the noise operator introduced by the classical amplification chain following the KIPA, while $a_\text{out}$ characterizes the KIPA's output. By utilizing the above equation, the total noise quanta of the entire amplification chain can be computed (see the Supplementary Materials for further details)
\begin{equation}
  N_\text{tot}\approx \hbar \omega G_\text{tot} \Big[\frac{1}{2}\,\text{coth}\Big(\frac{\hbar \omega}{2k_BT}\Big)+n_{\text{add}}\Big]
\end{equation}
where $G_\text{tot}=G_\text{h}G_\text{k}$ is the total gain and 
 \begin{align}\label{nadd1}
     n_{\text{add}}=\frac{G_\text{k}-1}{G_\text{k}}\Big(\bar n+\frac{1}{2}+n_{\text{k}}\Big)+\frac{n_{\text{h}}}{G_\text{k}},
 \end{align}
 is the total input-referred noise of the amplification chain. Here, $\bar{n}(T)=(e^{\hbar \omega/k_BT}-1)^{-1}$ is the thermal noise at temperature $T$, $k_B$ is the Boltzmann constant, and
\begin{equation}\label{noise-eq}
n_{\text{k}}=2\Big(\frac{1-\eta}{\eta}\Big)\Big[\bar{n}(T_\text{dev})+\frac{1}{2}\Big],
\end{equation}
is the input-referred noise added by the KIPA at a given device temperature $T_\text{dev}$. In perfect coupling regime $\eta\approx 1$, a very low temperature $\bar n \approx 0$, and at the large gain $G_{\text{k}}\gg 1$ such that $\frac{n_{\text{h}}}{G_\text{k}}\approx 0$, the total noise of the amplifier reduces to the vacuum noise $n_{\text{add}}\approx\frac{1}{2}$, as anticipated for an ideal non-degenerate quantum-limited amplifier.

By generating a known noise using a temperature-controlled $50\,\Omega$ load noise source and feeding it into the KIPA, we estimate the overall device noise at various gains, see  Fig. \ref{Fig6}a. First, during the noise calibration with the pump off, we deduce the gain $G_\text{h}$  and added noise  $n_\text{h}$ of the classical amplification chain. Next, with the pump turned on, we measure the system's noise to determine $G_\text{k}$ and $n_\text{add}$ while the system operates in the single-mode amplification regime, as shown in Fig. \ref{Fig6}b. In this figure for small KIPA gains ($G_\text k\approx0$), the noise from the classical amplification chain becomes the main source of noise i.e. $n_\text{add}=n_\text{h}$. Conversely, at significantly high gains, the KIPA noise dominates the overall system noise $n_\text{add}\approx (\bar n+\frac{1}{2}+n_{\text{k}})$. We use Eq. (\ref{nadd1}) to fit the experimental result, yielding $n_{\text{k}}\approx 0.82$ %at around $230$ mK, a close match to the theoretical estimation obtained from Eq. (\ref{noise-eq}) of $n_\text k\approx 0.39$ 
quanta noise. Part of the added noise comes from the rise in temperature of the mixing chamber of the dilution refrigerator, consequently increasing the sample's temperature to nearly 100 mK due to the application of DC-current to the sample. This issue can be addressed by adjusting the width of the sample and utilizing thinner wires.

We additionally note that here we employed the broadband variable temperature stage method to calibrate the noise characteristics of our KIPA. One challenge associated with this approach is that the calibration tool used generates parasitic heat, which perturbs the mixing chamber of the dilution refrigerator and affects the effective temperature of the sample. Nonetheless, in the Supplementary Materials, we have introduced an alternative approach to calibrate the sample noise. Other approaches, such as utilizing shot noise tunnel junctions, offer a potential avenue for more thorough validation of the added noise by the amplifier \cite{2312.14900}.

As a final remark, we describe the operational performance of our device at different temperatures. The kinetic inductance properties of KIPA, along with its independence from Josephson junctions, offer a substantial advantage when it comes to operating it at higher temperatures \cite{PhysRevApplied.17.044009}, a limitation that affects JPA and JTWP devices. We can test and operate the amplifier at different temperatures and measure the gain at different phases of the process. 
In Fig. \ref{Fig6}c, the gain of the amplifier in the single mode is illustrated at various device temperatures. It is apparent that the amplifier effectively maintains its gain performance, even at temperatures reaching up to 4.5 K. However, due to technical limitations and the inability to regulate the dilution refrigerator's temperature beyond 4.5 K, evaluating the device's performance at higher temperatures was unattainable.

\section{Conclusion and Discussions}\label{conclusion}
In summary, we have developed a junction-free quantum-limited amplifier based on kinetic inductance superconductivity. The design of our amplifier is inherently simple and does not necessitate complex circuitry or impedance-matching step couplers for amplification generation \cite{PhysRevApplied_jarrydPla}. Overcoming the limitations of conventional Josephson junction amplifiers, this device operates at temperatures above 4.5 K. Its double-mode capability, tunability through bias current, facilitates selective operation in both single and double-mode amplification regimes, achieving gains surpassing 50 dB in single-mode and 32 dB in double-mode configurations while only adding $0.82$ quanta noise. Compared to Josephson junction-based amplifiers our device presents a remarkably improved 1 dB compression point of $-73$ dB at $20$ dB gain mainly due to the small contribution of the self-Kerr term in the system. 
However, even though we did not measure the magnetic field dependency, the high-kinetic inductance NbTiN resonators demonstrate exceptional magnetic field compatibility, with fields up to 6 T \cite{PhysRevApplied.5.044004}. Moreover, recent studies affirm the maintenance of amplification performance up to 1 T \cite{PhysRevApplied.19.034024, 2311.07968, PhysRevApplied.21.024011}.

This amplifier can potentially be used in a broad range of quantum applications, particularly in the domain of quantum computing. Its adaptability and resilience make it a fitting candidate for integration into future superconducting quantum computers, enhancing microwave measurements and enabling fast and accurate readout of superconducting qubits and spins. Additionally, its ability to perform effectively at higher temperatures suggests its utility in quantum sensing and reading applications, such as quantum illumination and radar \cite{2310.07198}. Moreover, its resilience to magnetic fields opens possibilities for integration into hybrid quantum circuits or for the readout of spins and colored centers that necessitate magnetic fields for their control.

The low loss of our sample enables the generation of nonclassical radiation, such as entanglement and squeezing within the circuit. Quantum entanglement finds applications in quantum sensing, linking remote quantum nodes, and establishing entangled clusters across a chip. Squeezing, however, is a valuable resource for universal quantum computing \cite{PhysRevLett.97.110501, PhysRevLett.112.120504}. The simple design, high fabrication yield, minimal loss, and scalability of our device could potentially find applications in microwave continuous variable quantum computing \cite{Madsen2022}.

\textbf{Acknowledgments} We thank Joe Salfi, Leonid Belostotski, and Mohammad Khalifa
 helpful comments and discussions. S.B. acknowledges funding by the Natural Sciences and Engineering Research Council of Canada (NSERC) through its Discovery Grant and Quantum Alliance Grant, funding and advisory support provided by Alberta Innovates (AI) through the Accelerating Innovations into CarE (AICE) -- Concepts Program, support from Alberta Innovates and NSERC through Advance Grant project, and Alliance Quantum Consortium. This project is funded [in part] by the Government of Canada. Ce projet est financé [en partie] par le gouvernement du Canada.

\bibliography{AbdulReferences}

\clearpage

\onecolumngrid
\appendix
\begin{center}
\textbf{\large Supplementary Materials}
\end{center}

\subsection{Measurement setup}
\begin{figure}[h]
     \includegraphics[scale=1]{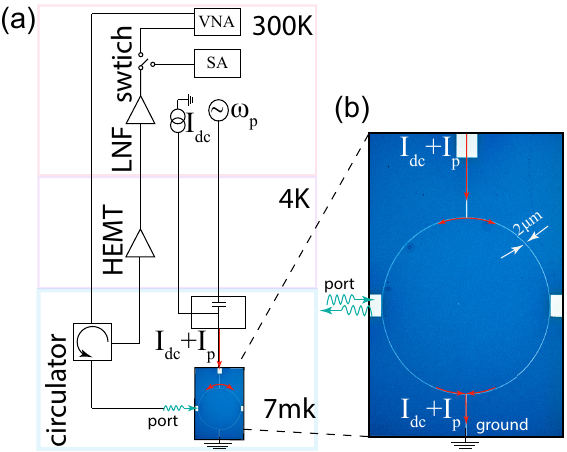}
        \caption{ (a) A schematic representation of the measurement setup for our device. A bias tee mixes the DC and pump current and is supplied to the KIPA galvanically. The attenuated input signal is sent to the device and then directed to a HEMT at 4K and room temperature amplifier (LNF) at 300K and is measured by either a spectrum analyzer (SA) or vector network analyzer (VNA). (b) The KIPA is a $1300\,\mu$m diameter ring resonator with a $2\,\mu$m  pitch, capacitively connected to an L-shaped resonator. The system is probed via a transmission line, capacitively connected to the left side of the ring. An additional line supplies the pump and DC current from the device's top, flowing to the ground through the ring.}
        \label{FigSI1}
\end{figure}

A bias tee is used to combine dc current $I_\text{dc}$ and pump $I_p$ at the mixing chamber of the dilution refrigerator, as seen in Fig. \ref{FigSI1}. The output of the bias tee is directly connected to KIPA. Applying a biased current leads to a shift in the resonance frequency, as explained in the main text. This shift allows us to infer $I^*$. Consequently, we select $I_\text{dc}$ to maximize the ratio $I_\text{dc}/I^*$ while remaining far from the sheet's critical current $I_\text{c}$. This allows having enough room to not break superconductivity by applying the pump to the system. By introducing a pump and sweeping its frequency, we can identify the optimal amplification point. This measurement is conducted using either a Vector Network Analyzer (VNA) or a Spectrum Analyzer (SA). 

For probe amplification, we utilized the VNA (Rohde and Schwarz ZBN20) and aligned the center frequency of the VNA trace to half of the pump frequency. By sweeping the pump frequency and measuring with the VNA, we can capture the maximum gain at various DC currents. The assessment of noise amplification was conducted using an SA (Rohde and Schwarz FSW), where the output power spectral density enables the measurement of vacuum noise amplification within the KIPA. This particular measurement allows for the determination of the noise added by the KIPA during the amplification process.

\subsection{Sample Fabrication and Packaging}

The fabrication process involves a straightforward single lithographic step facilitated by an optical lithographer. Initially, a high-resistivity intrinsic silicon ($R \geq 20\,k\, \Omega- \text{cm}$) wafer with a thickness of $500\, \mu m$ is coated with a $10$ nm layer of NbTiN and diced into smaller chips. 
The fabrication procedure starts with a cleaning step involving Acetone and IPA, where the chip is subjected to a 5-minute sonication treatment in each solution. The chip's surface is subsequently coated with AZ 1529 resist followed by the Optical Lithography step to pattern the design. The final step involves etching the device using an ICP Reactive Ion Etcher. Subsequently, the chip is secured using silver paste onto a gold-plated PCB. This PCB is fabricated from a Rogers AD 1000 laminate with a dielectric thickness of $0.5$ mm, and is coated with oxygen-free copper on both sides with a gold-plated finish. The PCB includes four $50 \,\Omega$ coplanar waveguide traces, which are connected to the external measurement line via a surface-mount mini-SMP microwave connector at one end. At the opposite end, the PCB is wire-bonded to the input and pump/DC ports of the KIPA resonator using aluminum bond wires. To address unwanted parasitic modes, an array of vias connects the transmission line on the PCB to both top and bottom ground planes. The entire PCB is then mounted inside an oxygen-free box, creating a 3D cavity with a resonance frequency far from the operational point of the sample. This 3D cavity shields the sample from parasitic noises and improves the thermalization of the sample.

\subsection{Sample Design}

The primary aim in designing our amplifier is to have a well-defined resonance that could be consistently measured and tracked while adjusting the frequency to targeted regions. To achieve this, we opted for a ring resonator that is capacitively coupled to a coplanar waveguide (CPW) transmission line. The ring is grounded to supply DC and pump current, effectively forming $\lambda$/4 resonator. While the aforementioned setup can enable single-mode amplification, we introduced an additional L-shaped resonator to generate double-mode amplification by capacitively coupling it to the ring using a coupling rate $J$.  

The ring resonator has a resonance frequency of $7.4$ GHz at zero dc-current. Its specific dimensions, including a radius of $650\,\mu$m and a track width of $2\,\mu$m, were determined through simulations conducted in Sonnet. The resonator's width was adjusted to minimize Kerr effects.

\begin{figure}[t]
     \includegraphics[width=1\textwidth]{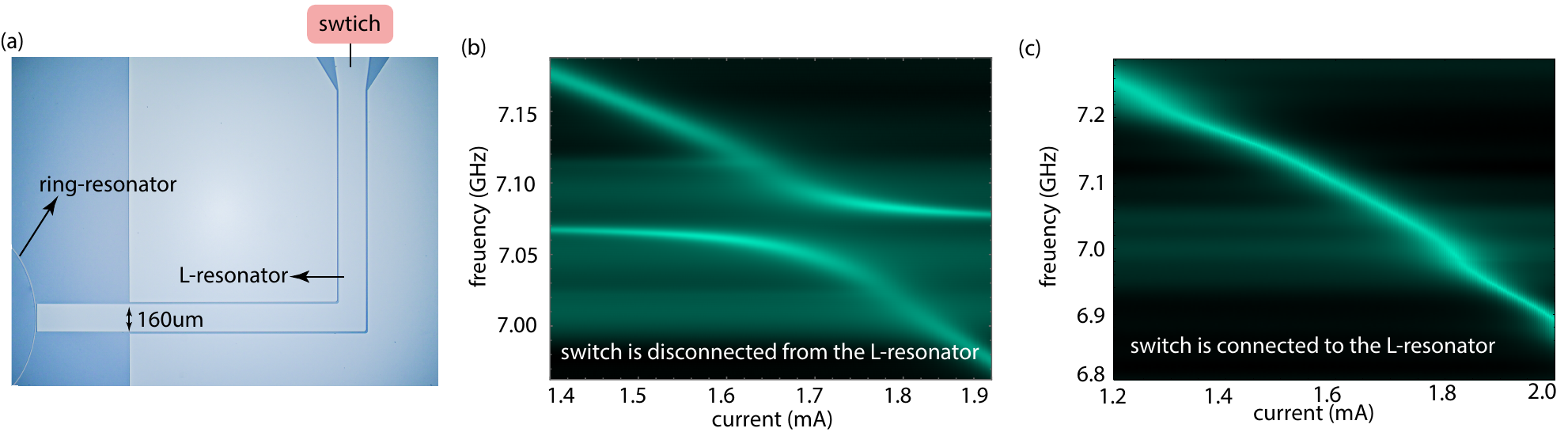}
       \caption{ (a) The optical image of the L-resonator capacitively coupled to the ring-resonator. The L-resonator is wirebonded to the PCB which is connected via an RF cable to a cryogenic microwave switch. (b) A close-up view of the anticrossing point after disconnecting the cryogenic switch from the L-shaped resonator. In this configuration, the anticrossing point becomes distinctly visible.  (c) The system's reflection is plotted against the applied dc-current in the presence of the cryogenic switch connected to the auxiliary resonator. This configuration results in a very large extrinsic coupling rate, causing the anticrossing point to be obscured. }
        \label{FigSI2}
\end{figure}

The auxiliary resonator is designed in an L-shaped structure with a specific width of $160\,\mu$m and length of $4$ mm, see Fig. \ref{FigSI2} a. Its considerable size offers two advantages. First, it facilitates direct wire bonding to the input port on the hosting PCB, serving as a port for measurements or probing within the system. By employing the cryogenic latching switch (Radiall R573423600) and connecting this resonator to it, we can even directly measure the system's reflection. Second, it can be considered as a linear resonator with a second resonance frequency closer to the ring resonator. 

An interesting aspect of this design is when the switch is disconnected from the L-resonator, allowing the resonator to float freely at both ends. This configuration demonstrates clear anticrossing behavior, attributed to its coupling with the ring resonator (see Figure \ref{FigSI2} b). The first resonance mode of this resonator occurs at around $4.4$ GHz, while the second resonance mode is at $7.03$ GHz. However, when the switch is connected to the L-shaped resonator, its extrinsic coupling $\kappa_\text e$ becomes significantly large, resulting in an extremely flattened mode that is not easily observable at the anticrossing points, as shown in Fig. \ref{FigSI2}c.

\subsection{System Noise calibration}\label{noisecal}
We determine the system gain $G$ and the system noise $n_{\text{add}}$ for both the HEMT and KIPA by introducing a known quantity of thermal noise using temperature-controlled $50\,\Omega$ load. By varying the temperature of the $50\,\Omega$ load and recording its temperature, we can generate predictable thermal noises. These noises will serve the purpose of calibrating the measurement chain. The calibration devices are connected to the measurement setup via two 5-cm-long superconducting coaxial cables and a thin copper braid (providing weak thermal anchoring to the mixing chamber plate) using a latching microwave switch (Radiall R573423600). By measuring the noise density (in square volts per hertz) at different temperatures and fitting the collected data with the anticipated scaling,
\begin{equation}
    N_\text{tot}=G_\text{tot}\hbar \omega \Big[\frac{1}{2}\text{coth}\Big(\frac{\hbar \omega}{2k_BT}\Big)+n_{\text{add}}\Big],
\end{equation}
we accurately determine the gain $G_\text{tot}$ and the added noise photons $n_{\text{add}}$ for each section of the amplification chain.

\subsection{Change in extrinsic and intrinsic couplings}
\begin{figure}[h]
     \includegraphics[scale=0.6]{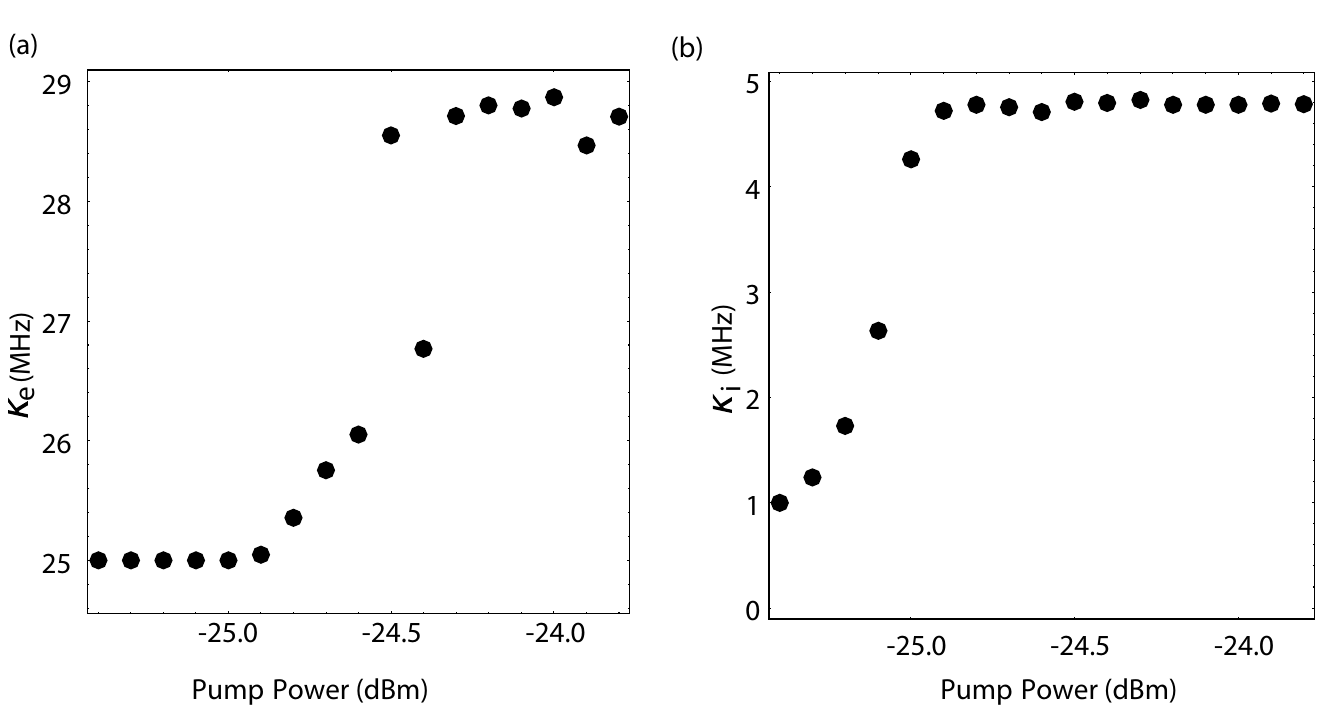}
       \caption{ (a) The extrinsic  $\kappa_e$ and (b) intrinsic coupling rate at different pump powers, extracted by fitting different amplification profiles.}
        \label{FigSI3}
\end{figure}

As previously discussed, the variability in both extrinsic and intrinsic coupling rates is anticipated due to the dependence of kinetic inductance on the RF current applied to the device. Figure \ref{FigSI3} illustrates the variations in both $\kappa_e$ and $\kappa_i$ for the ring-resonator as the pump power to the device is increased. The precise determination of pump current is challenging due to the unknown impedance of the device. Nonetheless, a noticeable trend emerges, showing that an increase in pump power corresponds to an increase in coupling rates. The rise in $\kappa_e$ is influenced by its dependence on inductance, typically a constant determined solely by the device's geometry. However, in our case, the kinetic inductance, which significantly is bigger than the geometric inductance, dominates the total inductance, impacting $\kappa_e$. The increase in $\kappa_i$ is attributed to impurities and defects in the device fabrication, leading to an increase in $\kappa_i$ when stimulated by current. 

\subsection{Gain-Bandwidth product}
\begin{figure}[h]
     \includegraphics[scale=0.6]{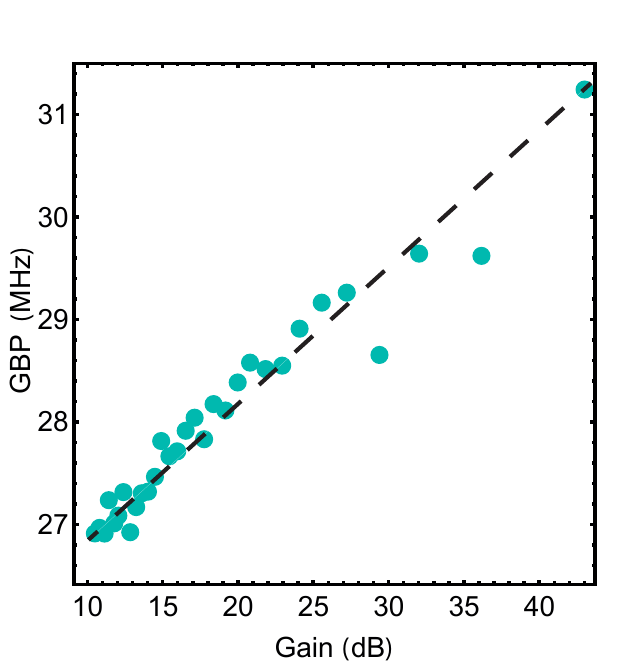}
       \caption{ Gain bandwidth product (GBP) at different gains of the KIPA.}
        \label{FigSI4}
\end{figure}

Subsequently, we proceed to extract the bandwidth and gain bandwidth product (GBP) at different gains. Bandwidth (BW) is defined as the full half-width maximum of linear gain, and GBP is defined as $\sqrt{G_\text{lin}}*BW$. Measured linear gain data is fitted to a Lorentzian line shape to extract the bandwidth. For sufficiently large gains, we anticipate GBP to be on the order of $\kappa_e$ as confirmed in Fig. \ref{FigSI4}.

\subsection{Kinetic Inductance}
The total kinetic inductance of both resonators, $L_0$, is determined by the sheet inductance of the thin film, which varies based on the film's thickness. We deduce the sheet inductance $L_{\square}$ by analyzing the resonance frequency of the ring resonator at zero bias current. Utilizing simulations in Sonnet and adjusting the sheet inductance to align with the experimental result, we can infer the sheet inductance of the film, resulting in the value of $L_{\square}=30$ pH. This analysis assumes a uniformly evaporated film across the entire chip. The total inductance corresponding to this sheet inductance then will be $L_0=251$ nH.

\subsection{Theory of the single-mode (degenerate) amplification}
The Hamiltonian describing a single mode resonator containing $\chi^{(2)}$ nonlinearity is given by
\begin{equation} \label{Hamsinglemode000}
    H=\omega_a a^\dagger a+\frac{g_0}{2} (a^2 a_p^\dagger+a^{\dagger2}a_p),
\end{equation}
where the second term describes the Three-Wave-Mixing (3WM) process, where a pump photon at frequency \(\omega_p\) is eliminated, producing two photons at the resonance frequency \(\omega_a\). Thus, the conservation of energy dictates \(\omega_p=2\omega_a\).

In the strong pump regime, we can ignore the field
fluctuations and the pump can be treated as a classical field with the amplitude $a_p\rightarrow \alpha_p=|\alpha_p|e^{-i\phi_p}e^{-i\omega_pt}$. A critical point to highlight is that if this approximation does not hold, our amplifier would not satisfy the linear amplification criteria we are aiming for. By using this approximation, we can subsequently simplify our analysis by neglecting the dynamics of the pump degree of freedom and focusing on the reduced system Hamiltonian
\begin{equation} \label{Hamsinglemode00}
    H=\Delta a^\dagger a+\frac{g}{2} (a^2+a^{\dagger2}),
\end{equation}
where $g=g_0|\alpha_p|$ and $\phi_p$ is the global phase set by the pump. The above Hamiltonian has been written in a reference frame rotating at $\omega_p/2=\omega_a-\Delta$.

The complete quantum description of the system can be expressed using the quantum Langevin equations. These equations incorporate quantum noise affecting input fluctuations for the resonator ($a_\mathrm{e}$ with extrinsic damping rate $\kappa_\mathrm{e}$) and the intrinsic losses in the resonator mode ($a_\mathrm{i}$ with intrinsic damping rate $\kappa_\mathrm{i}$). The correlations for these noises are defined as follows:
\begin{eqnarray}
    \langle a_{\mathrm{e(i)}} (t) a_{\mathrm{e(i)}}^\dagger (t')\rangle&=&\langle a_{\mathrm{e(i)}}^\dagger  (t) a_{\mathrm{e(i)}}(t')\rangle+\delta (t-t')\nonumber\\
    &=&(\bar{n}_{\mathrm{e(i)}}+1)\delta(t-t'),
\end{eqnarray}
where $\delta (t)$ is the Dirac delta function and $\bar{n}_{\mathrm{e(i)}}$ are the Planck-law thermal occupancies of microwave mode (bath). The resulting Langevin
equations corresponding to Hamiltonian (\ref{Hamsinglemode00}) are

\begin{eqnarray}
 \dot{a}&=&-\Big(i\Delta+\frac{\kappa}{2}\Big)a-ig e^{i\phi_p}a^\dagger+\sqrt{\kappa_\mathrm{e}}a_{\mathrm{e}}+\sqrt{\kappa_\mathrm{i}}a_{\mathrm{i}},\nonumber\\
  \dot{a}^\dagger&=&\Big(i\Delta-\frac{\kappa}{2}\Big)a^\dagger+ig e^{-i\phi_p}a+\sqrt{\kappa_\mathrm{e}}a_{\mathrm{e}}^\dagger+\sqrt{\kappa_\mathrm{i}}a_{\mathrm{i}}^\dagger,
     \label{eqmotionSI}
\end{eqnarray}
where $\kappa=\kappa_{\mathrm{e}}+\kappa_{\mathrm{i}}$ is the total damping rate of the resonator. We can solve the above equations in the Fourier domain to obtain the intra-cavity operator $a$. By substituting the solutions of Eqs. (\ref{eqmotionSI}) into the corresponding input-output relation i.e., $a_{\mathrm{out}}=\sqrt{\kappa_{\mathrm{e}}}a-a_{\mathrm{in}}$, we obtain
\begin{equation}
    \textbf{S}_\mathrm{out}(\omega)=\textbf{M} (\omega)\textbf{S}_\mathrm{in}(\omega)
\end{equation}
where $\textbf{S}_\mathrm{out}=[a_{\mathrm{out}},a_{\mathrm{out}}^\dagger]^\textbf{T}$,   $\textbf{M}(\omega)=\Big(\textbf{C}.[\textbf{A}]^{-1}.\textbf{B}-\textbf{D}\Big)$ with $\textbf{I}$ is the identity matrix, and $\textbf{S}_\mathrm{in}=[a_{\mathrm{e}},a_{\mathrm{i}},a_{\mathrm{e}}^\dagger,a_{\mathrm{i}}^\dagger]^\textbf{T}$, and we defined the following matrices
\begin{equation}
    \textbf A=-i\omega \textbf{I}-\begin{bmatrix}
-\Big(i\Delta+\frac{\kappa}{2}\Big) & -ig e^{i\phi_p} \\
ig e^{-i\phi_p} & \Big(i\Delta-\frac{\kappa}{2}\Big)
\end{bmatrix},
\end{equation}
\begin{equation}
    \textbf B=\begin{bmatrix}
\sqrt{\kappa_{\mathrm{e}}} & \sqrt{\kappa_{\mathrm{i}}} & 0 & 0 \\
 0 & 0&  \sqrt{\kappa_{\mathrm{e}}} & \sqrt{\kappa_{\mathrm{i}}}
\end{bmatrix},
    \textbf D=\begin{bmatrix}
1 & 0 & 0 & 0 \\
 0 & 0&  1 & 0
\end{bmatrix}
\end{equation}
and $\textbf{C}=\sqrt{\kappa_{\mathrm{e}}} \textbf{I}$. Note that the inverse of the matrix $\textbf{A}$ at $\Delta=0$ gives the susceptibility of the resonator $\boldsymbol{\chi}(\omega)=\textbf{A}^{-1}$.  
%\begin{equation}
%\Gamma (\omega) \equiv \frac{\frac{1}{2} \kappa_\mathrm{e} \kappa+i \kappa_\mathrm{e} \left(\Delta +\omega -\frac{\omega
  % _p}{2}\right)}{\Delta ^2+\left(\frac{\kappa }{2}+i \left(\omega -\frac{\omega _p}{2}\right)\right){}^2-|g|^2}-1,
  %  \label{eq:reflectionCavity}
%\end{equation} 
The total output field then has the following form
\begin{eqnarray}\label{aoutSI}
  a_\mathrm{out}(\omega)=\Big[\mathcal{G}_S(\omega)a_\text{e}+\mathcal{G}_I(\omega)a_\text{e}^\dagger\Big]+\sqrt{\frac{1-\eta}{\eta}}\Big[(\mathcal{G}_S(\omega)+1)a_\text{i}+\mathcal{G}_I(\omega)a_\text{i}^\dagger \Big]  
\end{eqnarray}
 where $\eta=\frac{\kappa_\text{e}}{\kappa}$ describes the waveguide-resonator coupling and we define
 \begin{eqnarray}
   \mathcal{G}_S(\omega)&=&\frac{\eta \kappa \Big[\frac{\kappa}{2}-i(\omega+\Delta)\Big]}{\Delta^2-g^2+(i\omega-\frac{\kappa}{2})^2}-1,\nonumber\\
   \mathcal{G}_I(\omega)&=&\frac{-i\eta \kappa g e^{i\phi}}{\Delta^2-g^2+(i\omega-\frac{\kappa}{2})^2}.
 \end{eqnarray}

The linear amplification model being analyzed here must hold the bosonic commutation relation for the output field operator $[  a_\mathrm{out} (\omega),  a_\mathrm{out}^\dagger (\omega')]=\delta(\omega-\omega')$, and thus
\begin{equation}\label{commu}
    |\mathcal{G}_I|^2 \Big(1+\frac{1-\eta}{\eta}\Big)=|\mathcal{G}_S|^2+\Big(\frac{1-\eta}{\eta}\Big)|\mathcal{G}_S+1|^2-1,
\end{equation}
where for a lossless resonator ($\eta=1$) we get $|\mathcal{G}_I|^2 = |\mathcal{G}_S|^2 -1$, as presented in the main text.

\subsection{Theory of the double-mode (non-degenerate) amplification}
In this section, we present a comprehensive model describing the theory of amplification for two coupled resonators, with one resonator containing $\chi^{(2)}$ nonlinearity. We employ two approaches: one based on the bare mode representation and the other based on the hybridized mode. The theory model based on the bare mode has been used to fit the experimental results in Fig. \ref{Fig5}c. However, the expressions for signal/idler gain are very complex to extract meaningful interpretations. In contrast, the theory model based on the hybridized mode offers a much clearer insight into the dynamics of the system and the interpretation of the physics.

\subsubsection{Theory model based on the bare modes}
The Hamiltonian describing the coupled-mode system, including the parametric processes and operating in the presence of a pump with a frequency $\omega_p$ is given by (with $\hbar=1$)

\begin{equation} \label{eq:hamiltonianTM}
    H=\omega_{a} a^\dagger a+\omega_{b} b^\dagger b+\omega_{p} a_p^\dagger a_p+J(a^\dagger b+b^\dagger a)+\frac{g_0}{2} (a^2a_p^\dagger+a^{\dagger2}a_p). 
\end{equation}
where $a$ and $b$ are the annihilation operators of each mode with frequencies $\omega_a$ and $\omega_b$, respectively, and with coherent mode coupling strength $J$. Here, we ignore the self-Kerr terms in each resonator. 

The quantum Langevin
equations in the strong pump regime and a reference frame rotating at $\omega_p/2=\Omega-\Delta$ are then given by
\begin{eqnarray}
 \dot{a}&=&-\Big(i\Delta+\frac{\kappa_a}{2}\Big)a-ig e^{i\phi_p}a^\dagger-iJb+\sqrt{\kappa_{a,e}}a_{e}+\sqrt{\kappa_{a,i}}a_{i},\nonumber\\
  \dot{b}&=&-\Big(i\Delta+\frac{\kappa_b}{2}\Big)b-iJa+\sqrt{\kappa_{b,e}}b_{e}+\sqrt{\kappa_{b,i}}b_{i}
     \label{eqmotionSITM}
\end{eqnarray}
where  $\kappa_j=\kappa_{j,\mathrm{e}}+\kappa_{j,\mathrm{i}}$ is the total damping rate of the resonator $j=a,b$ with extrinsic (intrinsic) damping rate $\kappa_{j,\mathrm{e(i)}}$. Here, we consider near the anti-crossing point where $\Omega=\omega_a=\omega_b$. Just as in the preceding section, we can solve these coupled equations and their conjugate parts,  $a^\dagger$, and $b^\dagger$, by moving to the Fourier domain. Integrating these solutions with input-output theory enables us to determine the output field operators for each mode

\begin{eqnarray}\label{aoutSI}
  a_\mathrm{out}(\omega)&=&\Big[\mathcal{G}_{a,S}(\omega)a_\text{e}+\mathcal{G}_{a,I}(\omega)b_\text{e}^\dagger\Big]+A_{\text{noise}},\\
   b_\mathrm{out}(\omega)&=&\Big[\mathcal{G}_{b,S}(\omega)b_\text{e}+\mathcal{G}_{b,I}(\omega)a_\text{e}^\dagger\Big]+B_{\text{noise} },
\end{eqnarray}
where $O_{\text{noise}}$ are the noise operators and
\begin{align}\label{gaintwo}
  \mathcal{G}_{a,S} (\omega)&= -1-\frac{2\kappa _{a,e} \left(2 i \Delta _b+\kappa _b-2 i \omega \right) \left(-J^2+\frac{1}{4} \left(2 \left(\Delta _a+\omega \right)+i \kappa _a\right) \left(2 \left(\Delta _b+\omega \right)+i \kappa _b\right)\right)}{\left(2 i \Delta _b+\kappa _b-2 i \omega \right)I (\omega)-J^2 \left(2 \left(\Delta _a+\omega \right)+i \kappa _a\right) \left(2 \left(\Delta _b+\omega \right)+i \kappa _b\right)+4J^4}\nonumber\\
  \mathcal{G}_{a,I}(\omega)&= \frac{2g J e^{i \phi } \sqrt{\kappa _{a,e}} \sqrt{\kappa _{b,e}} \left(2 i \Delta _b+\kappa _b-2 i \omega \right)}{\left(2 i \Delta _b+\kappa _b-2 i \omega \right)I (\omega)-J^2 \left(2 \left(\Delta _a+\omega \right)+i \kappa _a\right) \left(2 \left(\Delta _b+\omega \right)+i \kappa _b\right)+4J^4},
\end{align}
with
\begin{eqnarray}
 I(\omega)=g^2 \left(-\kappa _b+2 i \left(\Delta _b+\omega \right)\right)-\left(2 i \Delta _a+\kappa _a-2 i \omega \right) \left(-J^2+\frac{1}{4} \left(2 \left(\Delta _a+\omega \right)+i \kappa _a\right) \left(2 \left(\Delta _b+\omega \right)+i \kappa _b\right)\right),
\end{eqnarray}

Note that $\mathcal{G}_{b,S} (\omega)$ and $\mathcal{G}_{b,I} (\omega)$ can be found by $a\rightarrow b$ in Eqs. (\ref{gaintwo}).

\subsubsection{Theory model based on the hybridized modes}
 As explained in the main text, in the strong pump regime and a reference frame rotating at $\omega_p/2=\Omega-\Delta$, the Hamiltonian of the coupled system in the presence of nonlinearity reduces to
\begin{eqnarray} \label{eq:hamiltonianSI}
H=\Delta_+c_{+}^{\dagger}c_{+}+\Delta_- c_{-}^{\dagger}c_{-}+\frac{g}{2}\Big[e^{-i\phi_p}\big(c_{+}^{2}+c_{-}^{2}\big)+\text{h.c}\Big]+g\Big[e^{-i\phi_p}c_{+}c_{-}+\text{h.c}\Big]. 
\end{eqnarray}
where $g=g_0|\alpha_p|$,  $\Delta_{\pm}=\Delta\pm J$ and we consider $\Omega\equiv\omega_a=\omega_b$. 

By selecting $\Delta=J$ and employing the Rotating Wave Approximation (RWA), we effectively neglect the rapidly oscillating terms rotating at a frequency of $2J$. This decision allows us to select degenerate amplification in mode $c_-$, leading to the following equations of motion

\begin{eqnarray}
 \dot{c_-}&=&-\frac{\kappa_-}{2}c_- -ig e^{i\phi_p}c_-^\dagger+\sqrt{\kappa_-^\mathrm{e}}c_-^{\mathrm{e}}+\sqrt{\kappa_-^\mathrm{i}}c_-^{\mathrm{i}},\nonumber\\
  \dot{c_-}^\dagger&=&-\frac{\kappa_-}{2}c_-^{\dagger} +ig e^{-i\phi_p}c_-+\sqrt{\kappa_-^\mathrm{e}}c_-^{\mathrm{e}\dagger}+\sqrt{\kappa_-^\mathrm{i}}c_-^{\mathrm{i}\dagger},
     \label{eqmotionSIpm}
\end{eqnarray}
where $\kappa_{-}$ and $\kappa_{-}^{\text{e/i}}$ are the coupling rates of the collective operators with corresponding noise operators $c_-^{\mathrm{i/e}}$. We note that by selecting $\Delta=-J$ the $c_+$ dominates the amplification process. In this case, the equations of motion will be identical to Eqs. (\ref{eqmotionSIpm}) by replacing all $+\rightarrow -$. 

Transitioning to the Fourier domain and applying input-output theory allows us to determine the output operator for modes $c_\pm$
\begin{eqnarray}\label{aoutSI}
  c_\pm^\mathrm{out}(\omega)=\Big[\mathcal{G}_S^{\pm}(\omega)c_\pm^\text{e}+\mathcal{G}_I^{\pm}(\omega)c_\pm^{\text{e}\dagger}\Big]+\sqrt{\frac{1-\eta_\pm}{\eta_-}}\Big[(\mathcal{G}_S^{\pm}(\omega)+1)c_\pm^\text{i}+\mathcal{G}_I(\omega)c_\pm^{\text{i}\dagger} \Big]  
\end{eqnarray}
 where $\eta_\pm=\frac{\kappa_\pm^\text{e}}{\kappa_\pm}$ and we define
 \begin{eqnarray}
   \mathcal{G}_S^\pm(\omega)&=&\frac{\eta_\pm \kappa_\pm \Big[\frac{\kappa_\pm}{2}-i\omega\Big]}{-g^2+(i\omega-\frac{\kappa_\pm}{2})^2}-1,\nonumber\\
   \mathcal{G}_I^\pm(\omega)&=&\frac{-i\eta_\pm \kappa_\pm g e^{i\phi}}{-g^2+(i\omega-\frac{\kappa_\pm}{2})^2}.
 \end{eqnarray}

Finally, selecting $\Delta=0$ provides access to the non-degenerate amplification part of the Hamiltonian i.e. $c_-c_+$ term, where the idler and signal are spectrally separated. The equations of motion in this case are given by
\begin{eqnarray}
 \dot{c_-}&=&-\frac{\kappa_-}{2}c_- -ig e^{i\phi_p}c_+^\dagger+\sqrt{\kappa_-^\mathrm{e}}c_-^{\mathrm{e}}+\sqrt{\kappa_-^\mathrm{i}}c_-^{\mathrm{i}},\nonumber\\
  \dot{c_+}&=&-\frac{\kappa_+}{2}c_+ -ig e^{i\phi_p}c_-^{\dagger}+\sqrt{\kappa_+^\mathrm{e}}c_+^{\mathrm{e}\dagger}+\sqrt{\kappa_+^\mathrm{i}}c_+^{\mathrm{i}\dagger},
     \label{eqmotionSIpm1}
\end{eqnarray}

We can solve these coupled equations and their conjugate parts  $\dot{c}_\pm^\dagger$ by moving to the Fourier domain and using the input-output theory for the collective modes
\begin{eqnarray}
  c_\mathrm{\pm,out}(\omega)=\mathcal{G}^{\pm}_S(\omega)c_{\pm}^e+\mathcal{G}_I^{\pm}(\omega)c_{\mp}^{e\dagger}
  +\sqrt{\frac{1-\eta^{\pm}}{\eta^{\pm}}}\Big[(\mathcal{G}^{\pm}_S(\omega)+1)c_{\pm}^i+\mathcal{G}^{\pm}_I(\omega)c_{\mp}^{i\dagger} \Big].\nonumber 
\end{eqnarray}
where we define the gain factor 
\begin{eqnarray}
    \mathcal{G}^{\pm}_S(\omega)&=&\frac{\eta \kappa_{\pm} \Big[\frac{\kappa_{\mp}}{2}-i(\omega+\Delta_{\mp})\Big]}{[i(\omega-\Delta_+)-\frac{\kappa_+}{2}][i(\omega+\Delta_-)-\frac{\kappa_-}{2}]-g^2}-1, \nonumber\\
   \mathcal{G}_I^\pm(\omega)&=&\frac{-i\sqrt{\eta_-\eta_+} \kappa_\pm g e^{i\phi}}{[i(\omega-\Delta_+)-\frac{\kappa_+}{2}][i(\omega+\Delta_-)-\frac{\kappa_-}{2}]-g^2}.
\end{eqnarray}
as presented in the main text.

\color{black}
\subsection{Theory of the added noise }
In this section, we present the theoretical model of the added noise within the amplification chain. The input signal traverses through the KIPA and then undergoes propagation via transmission cables before being amplified by a HEMT at the 4 K stage and a subsequent amplifier at room temperature. Here, the gain of the KIPA is given by \(G_{\text{k}}=|\mathcal{G}_S|^2\), while \(G_{\text{h}}\) represents the collective noise of both the HEMT and the room-temperature amplifier, containing all losses occurring in the cables between the calibration tool and the Spectrum Analyzer at room temperature.

 The output of the KIPA when operating in the nondegenerate parametric regime is given by Eq. (\ref{aoutSI}). The signal then passes through the amplification chain leading to
\begin{equation}
a_\text{meas}=\sqrt{G_\text{h}}a_\text{out}+\sqrt{G_\text{h}-1}h^\dagger,
\end{equation}
where $h$ is the noise operator added by the amplification chain after the KIPA. The total noise quanta can be calculated using the measured spectrum of the system at a given bandwidth $\text{BW}$ 

\begin{equation}
    \frac{S_\text{meas}}{\text{BW}}=\hbar \omega \Big(\frac{\langle \Delta I_\text{meas}^2\rangle+\langle \Delta Q_\text{meas}^2\rangle}{2}\Big)
\end{equation}
where $I_\text{meas}=\frac{a_\text{meas}+a_\text{meas}}{\sqrt{2}}$ and $Q_\text{meas}=\frac{a_\text{meas}-a_\text{meas}}{i\sqrt{2}}$ are the quadrature operators. Therefore the total noise quanta is given by
\begin{eqnarray}\label{noisegeneral}
N_\text{tot}=\frac{S_\text{meas}}{\text{BW}}=\hbar \omega G_\text{h} \Big[|\mathcal{G}_S(\omega)|^2+|\mathcal{G}_I(\omega)|^2+\Big(\frac{1-\eta}{\eta}\Big)\Big(|\mathcal{G}_S(\omega)+1|^2+|\mathcal{G}_I(\omega)|^2 \Big)\Big]\Big[\bar n
(T)+\frac{1}{2}\Big]+\hbar \omega(G_\text{h}-1)n_\text{h},
\end{eqnarray}
where $\bar{n}(T)=(e^{\hbar \omega/k_B T}-1)^{-1}$ is thermal noise and $n_\text{h}=\langle h^\dagger h\rangle+\frac{1}{2}$ is the noise added by the amplification chain after the KIPA. Using Eq. (\ref{commu}) we can simplify further Eq. (\ref{noisegeneral}), resulting in
\begin{eqnarray}\label{KIPAnoise}
N_\text{tot}=\hbar \omega G_\text{h} \Big[2|\mathcal{G}_S(\omega)|^2+2\Big(\frac{1-\eta}{\eta}\Big)|\mathcal{G}_S(\omega)+1|^2-1\Big]\Big[\bar n
(T)+\frac{1}{2}\Big]+\hbar \omega(G_\text{h}-1)n_\text{h},
\end{eqnarray}

We can extract the noise introduced by KIPA by comparing it to the general noise expression for a nondegenerate quantum-limited amplifier
\begin{eqnarray}\label{idealnoise2}
N_\text{tot}=\hbar \omega G_\text{h} \Big[G_k(\bar n+\frac{1}{2})+(G_k-1)(\bar n
+\frac{1}{2}+n_k)\Big]+\hbar \omega(G_\text{h}-1)n_\text{h},
\end{eqnarray}
in the large gain regime $\frac{G_\text{h}-1}{G_\text{h}}\approx 1$ the above equation reduces to

\begin{equation}\label{idealAmp}
  N_\text{tot}\approx G_\text{tot}\hbar \omega \Big[\frac{1}{2}\,\text{coth}\big(\frac{\hbar \omega}{2k_BT}\big)+n_{\text{add}}\Big]
\end{equation}
 where $G_\text{tot}=G_\text{h}G_\text{k}$ is the total gain and 
 \begin{align}
     n_{\text{add}}=\frac{G_\text{k}-1}{G_\text{k}}\Big(\bar n+\frac{1}{2}+n_{\text{k}}\Big)+\frac{n_{\text{h}}}{G_\text{k}},
 \end{align}
is the total input-referred noise added by the amplification chain. By substituting Eq. (\ref{KIPAnoise}) into Eq.(\ref{idealnoise2}) and  assuming \(G_{\text{k}}=|\mathcal{G}_S|^2\)   we can find the noise contribution of the KIPA in nondegenerate mode
\begin{equation}
n_{\text{k}}=2\Big(\frac{1-\eta}{\eta}\Big)\frac{(\sqrt{G_\text{k}}+1)^2}{G_\text{k}-1}\Big[\bar{n}+\frac{1}{2}\Big],
\end{equation}
where we assume $\mathcal{G}_s$ is real. For large gain $G_\text{k}\gg 1$, the above expression reduces to 
\begin{equation}
n_{\text{k}}=2\Big(\frac{1-\eta}{\eta}\Big)\Big[\bar{n}+\frac{1}{2}\Big],
\end{equation}
as presented in the main text. It is obvious that for a near ideal resonator $\eta\rightarrow 1$, the total noise of the amplification chain reduces to the noise expression for an ideal non-degenerate amplifier (for $G_\text k\gg 1$)

 \begin{align}
     n_{\text{add}}\approx\bar n+\frac{1}{2}.
 \end{align}

{
\subsection{Alternative approach for Noise calibration: Pump on/off measurement} 

The noise calibration technique discussed in Section \ref{noisecal} is inherently slow. Moreover, as the temperature of the calibration tool surpasses 1 K, it affects the temperature of both the mixing plate of the dilution refrigerator and subsequently the sample, potentially resulting in inaccurate outcomes. In this section, we present an alternative approach for noise calibration that is considerably faster and capable of determining the amplifier's added noise even at higher temperatures.  

The measurement setup is shown in Fig. \ref{FigSI5}a, which incorporates a well-known thermal noise source as a calibration reference for the amplification chain. This method is similar to the one described in Section \ref{noisecal}. The calibration tool is positioned in the output line of the amplification chain via a cryogenic switch, permitting easy switching between the calibration tool and KIPA as necessary. The total noise operator at the end of the measurement chain is given by 
\begin{equation}
  a_\text{meas}=\sqrt{G_\text{h}} \Big( \sqrt{\alpha} a_\text{out}
+\sqrt{(1-\alpha)}a_\text{env}\Big)+\sqrt{G_\text{h}-1}h^\dagger,
\end{equation}
where $a_\text{out}$ represents the output of the KIPA, and the parameter $\alpha$ shows the loss occurring between the sample and the switch which adds the noise operator $a_\text{env}$ to the measurement chain. We use a very short superconducting cable between the switch and the calibration tool so that we can safely neglect any significant loss. All other losses within the cables, along with their associated noise contributions after the switch, have been factored into the overall gain $G_\text{h}$ and noise operator $h$. The KIPA's gain can be accurately determined through the use of coherent tone amplification. The loss parameter $\alpha$ is also measurable by employing a VNA or SA. To address the remaining unknown parameters, a two-step noise calibration procedure is implemented.
In step 1, the calibration tool is used to assess the total gain ($G_\text{h}$) and noise properties ($n_\text{h}$) of the measurement setup without the involvement of the KIPA. In this setup, the KIPA is disengaged from the amplification chain. The complete symmetric measured spectrum is then given by
\begin{equation}
  \frac{S_\text{off}}{\text{BW}}=\hbar \omega G_\text{h} \Big[ \bar n (T)+\frac{1}{2}+\frac{(G_\text{h}-1)}{G_\text{h}}n_\text{h}\Big],
\end{equation}
where $\text{BW}$ is the measurement bandwidth,  $n_\text{h}=\langle h^\dagger h\rangle+1/2$ is the noise added by the amplification chain in the absence of the KIPA, and $\bar n(T)=(e^{\hbar \omega/k_bT}-1)^{-1}$. 

Step 2 involves switching to the KIPA configuration and assessing the gain through coherent tone amplification, along with noise amplification measurements performed at various temperatures. The total spectrum of the system in the presence of the KIPA is given by
\begin{figure}[t]     \includegraphics[width=1\linewidth]{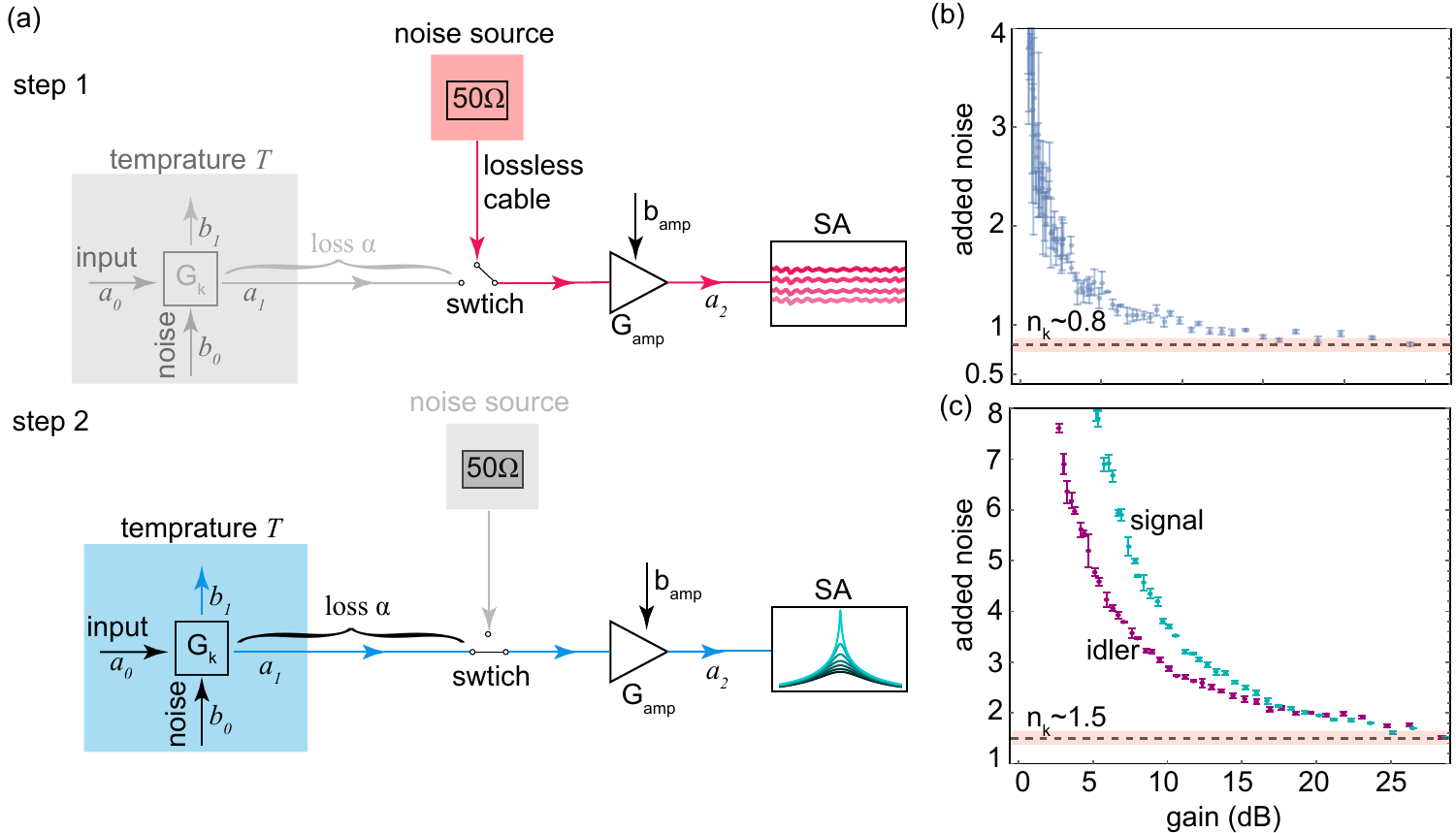}
       \caption{ (a) The configuration utilized for assessing the noise characteristics of the KIPA. In step 1, a switch enables the calibration of the total gain and noise within the classical amplification chain, shown by $G_k$ and $n_k$ respectively. In step 2, switching to KIPA allows for the precise determination of the gain and noise introduced by the quantum-limited amplifier. The noise introduced by the KIPA versus the gain of the KIPA for (b) single-mode and (c) two-mode amplification regimes. }
        \label{FigSI5}
\end{figure}
\begin{eqnarray}
  \frac{S_\text{on}}{\text{BW}}=\hbar \omega G_\text{h} \Big[ \alpha G_\text k\Big(\bar n+\frac{1}{2}+n_\text{add}\Big)+(1-\alpha)(\bar n+\frac{1}{2})\Big],\nonumber\\
\end{eqnarray}
where $  n_{\text{add}}=\frac{G_\text{k}-1}{G_\text{k}}\Big(\bar n+\frac{1}{2}+n_{\text{k}}\Big)+\frac{n_{\text{h}}}{G_\text{k}}$ is the total noise added by the amplification chain.
By measuring $G_\text{k}$ from coherent probe amplification and $G_\text{h}$ from step 1, we can infer the total noise added by the KIPA
\begin{equation}\label{noiseamp}
  n_\text{k}\approx\Big[\frac{S_\text{on}-S_\text{off}}{\hbar \omega\text{BW}(G_\text{k}-1)G_\text{h}\alpha}-\Big(2\bar n+1\Big)\Big],
\end{equation}

The measurement results for single-mode and two-mode amplification are shown in Fig. \ref{FigSI5}b and c, respectively. The noise calibration for single mode yields approximately $0.82$ confirming the result from the noise calibration approach used in Section \ref{noisecal}. Operating the system in two-mode amplification results in approximately $1.5$ quanta of noise added by KIPA, as seen in Fig. \ref{FigSI5}c.}

\end{document}